\documentclass[review]{elsarticle}
\usepackage{geometry}
\journal{European Journal of Operational Research}
\usepackage{amssymb}
\usepackage{amsmath}
\usepackage{eucal}
\usepackage{color}
\usepackage[algo2e,ruled,vlined,]{algorithm2e}
\usepackage{algorithmic}
\usepackage[section]{placeins}
\usepackage[mathlines]{lineno}
\usepackage{mathtools}
\usepackage{setspace}
\usepackage{tabularx}
\usepackage{booktabs,multirow,array,multirow,graphicx}

\usepackage{float}

\usepackage{hyperref}
\newtheorem{theorem}{Theorem}
\newtheorem{lemma}[theorem]{Lemma}
\newtheorem{proposition}[theorem]{Proposition}

\newtheorem{Model}{Model}
\newproof{Proof}{Proof}
\newproof{pot}{Proof of Theorem \ref{thm2}}
\newfloat{Model_float}{htb}

\usepackage{color}
\definecolor{red}{rgb}{1,0,0}

\def\VRHDW#1#2#3{\vrule height #1 depth #2 width #3}
\newcommand{\up}{\VRHDW{1.5em}{0em}{0em}}
\newcommand{\down}{\VRHDW{0em}{0.625em}{0em}}%

\definecolor{green}{rgb}{0,.6,0}

\newcommand{\colored}{\operatorname{colored}}
\newcommand{\co}{\operatorname{count}}
\newcommand{\zeros}{\operatorname{zeros}}

\usepackage{siunitx}
\usepackage{etoolbox}
\newcommand{\ubold}{\fontseries{b}\selectfont}
\robustify\ubold

\begin{document}

\begin{frontmatter}
\title{Computational Approaches for Zero Forcing \\and Related Problems}
\author[Rice]{Boris Brimkov}
\ead{boris.brimkov@rice.edu}

\author[Rice]{Caleb C. Fast\corref{CCFemail}}
\address[Rice]{Department of Computational and Applied Mathematics, Rice University, 6100 Main St. - MS-134, Houston, Texas 77005}

\cortext[CCFemail]{Corresponding author.  Now at Business Analytics and Operations Research, FedEx Express, 3680 Hacks Cross Rd. - Building H 2nd Floor, Memphis, TN 38125}
\ead{calebfast@gmail.com}

\author[Rice]{Illya V. Hicks}

\ead{ivhicks@rice.edu}

\begin{abstract}
In this paper, we propose computational approaches for the zero forcing problem, the connected zero forcing problem, and the problem of forcing a graph within a specified number of timesteps. Our approaches are based on a combination of integer programming models and combinatorial algorithms, and include formulations for zero forcing as a dynamic process, and as a set-covering problem. We explore several solution strategies for these models, test them on various types of graphs, and show that they are competitive with the state-of-the-art algorithm for zero forcing. Our proposed algorithms for connected zero forcing and for controlling the number of zero forcing timesteps are the first general-purpose computational methods for these problems, and are superior to brute force computation.

\end{abstract}

\begin{keyword}
 Combinatorial optimization \sep zero forcing\sep integer programming\sep set-covering

\end{keyword}

\end{frontmatter}



\section{Introduction}
Zero forcing is an iterative graph coloring process where at each discrete timestep, a colored vertex with a single uncolored neighbor forces that neighbor to become colored. A zero forcing set of a graph is a set of initially colored vertices which forces the entire graph to become colored. The zero forcing number is the cardinality of the smallest zero forcing set. Zero forcing was initially introduced to bound the maximum nullity of the family of symmetric matrices described by a graph \cite{AIM-Workshop}; it was also independently studied in quantum physics \cite{quantum1} and theoretical computer science \cite{fast_mixed_search}, and has since found a variety of uses in physics, logic circuits, coding theory, power network monitoring, and in modeling the spread of diseases and information in social networks; see \cite{zf_tw,quantum1,logic1,powerdom3,proptime1,zf_np,powerdom2} and the bibliographies therein. 

Connected zero forcing is a variant of zero forcing in which the initially colored set of vertices induces a connected subgraph. The connected zero forcing number of a graph is the cardinality of the smallest connected set of initially colored vertices which forces the entire graph to be colored (i.e., the smallest connected zero forcing set). Applications and various structural and computational aspects of connected zero forcing have been investigated in \cite{brimkov2,brimkov,brimkov3}; in particular, it can be used for modeling the spread of ideas or diseases originating from a single connected source in a network, or for power network monitoring accounting for the cost of supporting infrastructure. 
Other variants of zero forcing, such as positive semidefinite zero forcing \cite{Barioli,positive_zf2,fractional_zf,proptime2}, fractional zero forcing, signed zero forcing \cite{signed_zf}, and $k$-forcing \cite{kforcing1,kforcing2} have also been studied. These are typically obtained by modifying the zero forcing color change rule, or adding certain restrictions to a zero forcing set. The number of timesteps in the zero forcing process after which a graph becomes colored is also a problem of interest (see, e.g., \cite{proptime_oriented,throttling,iteration_index,proptime1,proptime2}). Connected variants of other graph problems -- such as connected domination and connected power domination \cite{Caro,Desormeaux,FanWatson,connected_dom} -- have been extensively studied as well.

A closely related problem to zero forcing is power domination, where given a set $S$ of initially colored vertices, the zero forcing color change rule is applied to $N[S]$ instead of to $S$. Integer programming formulations for power domination and its variants have been explored in \cite{aazami_pd,brimkov_pd}. The power domination problem is derived from the phase measurement unit (PMU) placement problem in electrical engineering, which has also been studied extensively; see, e.g., \cite{pmu1,pmu2} and the bibliographies therein for various integer programming models and combinatorial algorithms for the PMU placement problem. Another closely related problem to zero forcing is the target set selection problem, where given a set $S$ of initially colored vertices and a threshold function $\theta: V(G) \rightarrow\mathbb{Z}$, all uncolored vertices $v$ that have at least $\theta(v)$ colored neighbors become colored. Thus, the zero forcing problem constrains the infectors, but the target set selection problem constrains the infectees.  See \cite{target1,target3,target2} for computational approaches of finding the smallest target set $S$ of initially colored vertices which causes the entire graph to become colored.

Computing the zero forcing number and connected zero forcing number of a graph are both NP-complete problems \cite{aazami,brimkov2,fast_mixed_search}; nevertheless, it is important to develop practical algorithms for solving these problems, at least on moderately-sized graphs. The state-of-the-art approach for computing the zero forcing number of a graph is a combinatorial algorithm called Wavefront, developed by Butler et al. \cite{WavefrontAlgorithm} (a version of this algorithm, altered for a related problem, appears in Butler et al. \cite{butler}). While this algorithm is the best available for the zero forcing problem, it is not flexible and cannot accommodate additional constraints, such as assuring connectivity of the solution or limiting the number of timesteps used to force the graph. 
A lot of effort has been put into developing closed formulas, efficient algorithms, characterizations, and bounds for the zero forcing numbers of graphs with special structure (see, e.g., \cite{AIM-Workshop,benson,brimkov2,brimkov,Edholm,Eroh,Huang,Meyer}), but relatively little progress has been made on developing computational methods for general graphs.

\subsection{Main Contributions}

In this paper, we explore approaches for computing the zero forcing number of a graph using integer programming. In particular, we present formulations of zero forcing based on two different perspectives -- one as a dynamic process, and the other as a set-covering problem. We explore several solution strategies of these models -- such as direct computation, constraint generation, and generation of facet-inducing constraints -- and compare their performance to Wavefront on different types of graphs.

We also propose a combinatorial algorithm for computing the connected zero forcing number of a graph, and extend the proposed integer programming models to the connected zero forcing problem by adding connectivity constraints. In doing so, we explore several different types of connectivity constraints which have been used in problems like connected domination, Steiner trees, and forest planning. Until now, there have not been any computational approaches for connected zero forcing of general graphs other than brute-force computation.

Finally, we adapt one of our integer programs to find zero forcing sets which force the graph within a specified number of timesteps, and have minimum cardinality among all such sets. To our knowledge, there have not been any previously-implemented algorithms for this problem (though some models have been proposed for the related problem of power domination \cite{aazami_pd}).

Our computational experiments show that our integer programming models are generally comparable to the Wavefront algorithm in sparse random graphs, and are superior to Wavefront in graphs corresponding to electrical power grids and other standard benchmark networks. Our proposed integer programming models for connected zero forcing significantly outperformed the combinatorial brute force and branch-and-bound algorithms. Moreover, in some cases, these approaches were faster, and able to handle larger graphs, than the Wavefront algorithm and the zero forcing analogues of the integer programs. This is somewhat surprising, since the connected variants of problems like domination and power domination have typically proven more difficult to solve computationally, due to their non-locality (see \cite{connected_dom} for more details). Since the connected zero forcing number is an upper bound to the zero forcing number, the proposed approaches for connected zero forcing can be used to obtain upper bounds or approximations to the zero forcing number, especially for graphs which are too large for Wavefront.

The paper is organized as follows. In the next section, we recall some graph theoretic notions, specifically those related to zero forcing. In Section \ref{section_comb_algorithms}, we present combinatorial approaches for computing the zero forcing number and connected zero forcing number of a graph; in Section \ref{sec:IPMethods}, we present integer programming approaches for these problems. In Section \ref{sec:Zero_Forcing_Results}, we describe the implementation of our proposed approaches, and compare them through computational experiments on various types of graphs. We conclude with some final remarks and open questions in Section~\ref{sec:ZeroComputationConclusions}.

\section{Preliminaries}
\label{sec:Preliminaries}

A graph $G=(V,E)$ consists of a vertex set $V$ and an edge set $E$ of two-element subsets of $V$.  In this paper, we consider \emph{simple} graphs, for which a subset $\lbrace v,w \rbrace \in E$ must have $v \neq w$ and $E$ contains at most one copy of $\lbrace v,w \rbrace$.  The \emph{order} and \emph{size} of $G$ are denoted by $n=|V|$ and $m=|E|$, respectively. Two vertices $v,w\in V$ are \emph{adjacent}, or \emph{neighbors}, if $\{v,w\}\in E$. The \emph{neighborhood} of $v\in V$ is the set of all vertices which are adjacent to $v$, denoted $N(v)$; the \emph{closed neighborhood} of $v$, denoted $N[v]$, is the set $N(v)\cup\{v\}$. Similarly, given $S\subset V$, $N(S)$ denotes the set $(\bigcup_{v\in S} N(v))\backslash S$, and $N[S]$ denotes the set $N(S)\cup S$. The \emph{degree} of $v\in V$ is defined as $d(v)=|N(v)|$. Given $S \subset V$, the \emph{induced subgraph} $G[S]$ is the subgraph of $G$ whose vertex set is $S$ and whose edge set consists of all edges of $G$ which have both endpoints in $S$. For other graph theoretic terminology and definitions, we refer the reader to~\cite{west}.

\begin{figure}[h!]
\begin{center}
\includegraphics[scale=0.3]{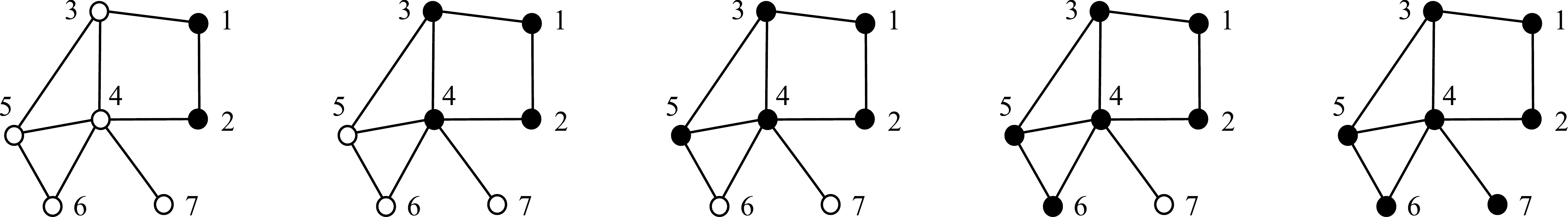}\qquad \qquad
\caption[Zero forcing in a graph]{\emph{Left:} A minimum zero forcing set of the graph is marked by colored vertices. Then, from left to right, the following forces are applied: $1\rightarrow 3$ and $2\rightarrow 4$; $3\rightarrow 5$; $5\rightarrow 6$; $4\rightarrow 7$.}
\label{fig_example_zf}
\end{center}
\end{figure}

Given a graph $G=(V,E)$ and a set $S \subset V$ of initially colored vertices, the \emph{color change rule} dictates that at each integer-valued timestep, a colored vertex $u$ with a single uncolored neighbor $v$ \emph{forces} that neighbor to become colored. 
The  \emph{closure} of $S$, denoted $\emph{cl}(S)$, is the set of colored vertices obtained after the color change rule is applied until no new vertex can be forced; it can be shown that the closure of $S$ is uniquely determined by $S$ (see \cite{AIM-Workshop}). A \emph{zero forcing set} is a set whose closure is all of $V$; the \emph{zero forcing number} of $G$, denoted $Z(G)$, is the minimum cardinality of a zero forcing set.  
A \emph{chronological list of forces} associated with a zero forcing set $Z$ is a sequence of forces applied to obtain the closure of $Z$ in the order they are applied. A \emph{forcing chain} for a chronological list of forces is a maximal sequence of vertices $(v_1,\ldots,v_k)$ such that $v_i$ forces $v_{i+1}$ for $1\leq i\leq k-1$. Each forcing chain is a distinct path in $G$, one of whose endpoints is an initially colored vertex; the other is called a \emph{terminal vertex}. See Figure \ref{fig_example_zf} for an illustration. A \emph{fort}, defined by Fast and Hicks \cite{ForcingTheoryPaper}, is a non-empty set $F\subset V$ such that no vertex outside $F$ is adjacent to exactly one vertex in $F$. In Figure \ref{fig_example_zf}, the sets $\{1,4,5,7\}$ and $\{2,3,6\}$ are forts. A zero forcing set \emph{restrained} by $S\subset V$ is a zero forcing set which contains $S$; $Z(G;S)$ denotes the cardinality of the smallest zero forcing set restrained by $S$ (cf. \cite{brimkov3}). 

A \emph{connected zero forcing set} of $G$ is a zero forcing set of $G$ which induces a connected subgraph. The \emph{connected zero forcing number} of $G$, denoted $Z_c(G)$, is the cardinality of a minimum connected zero forcing set of $G$. For short, we may refer to these as \emph{connected forcing set} and \emph{connected forcing number}. Note that a disconnected graph cannot have a connected forcing set. 

An \emph{empty} graph is a graph with no edges. A graph is \emph{cubic} if all of its vertices have degree 3. Let $C(n,k)$ be the graph with vertex set $\{0,\ldots,n-1\}$ and edge set $\{\{i,j\}:0\leq i,j\leq n-1,|i-j|\leq k/2\}$. A\emph{Watts-Strogatz} graph with parameters $n$, $k$, and $\beta$ refers to a graph obtained from $C(n,k)$ by replacing each edge of $C(n,k)$ with probability $\beta$ by a randomly chosen edge. When $n$ is clear from the context (or for a predetermined set of values of $n$), we will refer to the Watts-Strogatz graphs with parameters $n$, $k$, and $\beta$ as WS$(k,\beta)$. Watts-Strogatz graphs were introduced in \cite{WattsStrogatz}, and are a popular random graph model; they are meant to have small-world properties such as short average path lengths and high clustering.
Finally, we will use the notation $[n]$ to represent the set $\{1,\ldots,n\}$.

\section{Combinatorial Approaches}
\label{section_comb_algorithms}

In this section, we describe several combinatorial approaches for computing the zero forcing and connected forcing numbers of a graph $G=(V,E)$. The trivial, or \emph{brute force}, approach for finding a minimum zero forcing set of $G$ is to iteratively compute the closures of all subsets of $V$ of size $i$, starting from $i=1$ and incrementing $i$, until a zero forcing set is found. Similarly, to find a minimum connected forcing set, one could again generate subsets of vertices of increasing size, check whether each set induces a connected subgraph, and stop when the first connected set whose closure is $V$ is found. The closure of a set of vertices can be found in $O(m+n)$ time using Algorithm \ref{alg_closure}. 

\LinesNumbered
\begin{algorithm2e}[h]
\KwData{$G = (V,E)$ and a set $Z\subset V$}
\KwResult{The closure $cl(Z)$}
  $\colored \leftarrow\zeros(n)$\;
  $\co \leftarrow \zeros(n)$\;

\textbf{for} $v\in Z$ \textbf{do} $\colored(v)\leftarrow 1$\;

  \For{$v \in Z$}{
  $\co(v)\leftarrow \sum_{y\in N(v)}\colored(y)$\;

  \textbf{if} $\co(v)=d(v)-1$ \textbf{then}   push$(v)$\;

  }  
  \While{\emph{Stack is not empty}}{
  pop$(u)$\;
  $v\leftarrow y\in N(u):\colored(y)=0$\;  
  $\colored(v)\leftarrow 1$\;
  \For{$w\in N(v)$}{
  \If{$\colored(w)=1$}{
  	$\co(w)\leftarrow \co(w)+1$\;

\textbf{if} $\co(w)=d(w)-1$ \textbf{then}   push$(w)$\;

  }
  }
  $\co(v)\leftarrow \sum_{y\in N(v)}\colored(y)$\;

\textbf{if} $\co(v)=d(v)-1$ \textbf{then}   push$(v)$\;
  }
\Return $\{v:\colored(v)=1\}$\;
\caption{Finding the closure of a set in $O(n+m)$ time}
\label{alg_closure}
\end{algorithm2e}
%

\begin{proposition}
Let $G=(V,E)$ be a graph and $Z\subset V$. Algorithm  \ref{alg_closure} finds $cl(Z)$ in $O(m+n)$ time.
\end{proposition}
\begin{Proof}
Algorithm \ref{alg_closure} maintains an array ``$\colored$" which indicates whether a vertex is colored, an array ``$\co$" which counts the number of colored neighbors a vertex has, and a Stack containing \emph{active vertices}, i.e., colored  vertices which have a single uncolored neighbor. After the first two for-loops, the Stack contains all active vertices. In each iteration of the while-loop, an active vertex $u$ forces its uncolored neighbor $v$ and is removed from the Stack. The only inactive vertices which may become active as a result of $v$ becoming colored are the vertices in $N[v]$; all of these vertices are checked in the while-loop, and any active vertices among them are added to the Stack. Since one vertex is removed from the Stack (and can never re-enter the Stack) in each iteration of the while-loop, Algorithm \ref{alg_closure} terminates. When the Stack is empty, there are no more active vertices; thus, no more forces are possible, and the set of colored vertices at the end of the while-loop is exactly $cl(Z)$, as desired.

Lines 1---6 can be executed in $O(n+m)$ time since the neighborhood of each vertex is considered at most once. 
For each vertex $x$ that enters and exits the Stack, $O(d(x))$ operations are performed at most twice: once when $x$ is the unique uncolored neighbor of some other vertex (i.e., when $x=v$ in lines 11---16), and the second time when searching for the unique uncolored neighbor of $x$ (i.e., when $x=u$ in line 9). Thus, since each vertex enters and exits the Stack at most once, the total runtime of Algorithm \ref{alg_closure} is $O(m+n)+O(\sum_{x\in V}d(x))=O(n+m)$.
\hfill$\Box$
\end{Proof}

The brute force approach works well when the graph is known \emph{a priori} to have a very small or very large forcing number (in the latter case, one would start from $i=n$, decrement $i$ as soon as a forcing set is found, and stop when all sets of vertices of a certain size are not forcing). Similarly, the brute force approach can be used in conjunction with theoretical bounds on the forcing number in terms of other efficiently-computable parameters (see, e.g., \cite{AIM-Workshop,benson,brimkov2,brimkov,row}). In particular, if it is determined that $k_1\leq Z_c(G)\leq k_2<\frac{n}{2}$, it can be checked whether each of the $\binom{n}{k_1}+\cdots+\binom{n}{k_2}$ sets of vertices of appropriate size is connected and forcing in $O(m+n)$ time, so $Z_c(G)$ can be computed in $O((k_2-k_1)n^{2+k_2})$ time (the same applies to $Z(G)$). Other advantages of the brute force approach are that it is easy to implement, uses little memory, and can be easily parallelized, since closures of different sets of vertices can be computed independently.

Nevertheless, in practice, the brute force algorithm is usually outperformed by the other algorithms discussed in the sequel. Section \ref{sec:Wavefront} describes the Wavefront algorithm -- a dynamic programming style improvement of the brute force algorithm, which stores minimum forcing sets of certain subgraphs of $G$ and uses them to build minimum forcing sets of larger subgraphs. Thus, it avoids checking all possible subsets of vertices at the expense of increased memory. Section~\ref{sec:branch and bound} gives a branch-and-bound style improvement of the brute force algorithm for connected forcing; instead of generating all subsets of vertices and checking whether they are connected and forcing, this algorithm generates only connected subgraphs, checks whether they are forcing, and prunes the search tree based on the best zero forcing set found. 

\subsection{Wavefront Algorithm}
\label{sec:Wavefront}

In this section, we give a description of the combinatorial algorithm for zero forcing known as Wavefront, developed by Butler et al. \cite{WavefrontAlgorithm}. To our knowledge, this algorithm is the only previously-implemented computational method for the zero forcing problem (aside from brute force), and a proof of its correctness does not appear elsewhere in print. We prove that the Wavefront algorithm is correct in Theorem 
\ref{thm:WavefrontCorrectness} and give a result about its worst-case memory requirements in Theorem \ref{thm:WavefrontComplexity}.

\LinesNumbered
\begin{algorithm2e}[h]
\label{algorithm_wavefront}
\KwData{Graph $G = (V,E)$}
\KwResult{Zero forcing number of $G$}
$\mathcal{C} \leftarrow \lbrace (\emptyset, 0) \rbrace$\;
  \For{$R \in [n]$}{
  \For{$(S,r) \in \mathcal{C}$}{
  \For{$v \in V$}{

  $S' \leftarrow \emph{cl}(S \cup N[v])$\;    
  $r'\leftarrow r+|\{v\}\backslash S|+\max\{|N(v) \backslash S|-1,0\}$\;  

    \If{$r'\leq R$ \textbf{\emph{and}} $(S',i) \notin \mathcal{C}$ \textbf{\emph{for}} $i \leq R$}{
      $\mathcal{C}\leftarrow \mathcal{C} \cup \{(S',r')\}$\;
      \textbf{if} $S'=V$ \textbf{then return} $r'$\;

  }
  }
  }
  }

\caption{Wavefront Algorithm~\cite{WavefrontAlgorithm}\label{alg:Wavefront}}
\end{algorithm2e}
\LinesNotNumbered


\begin{lemma}
\label{lem:smallestsubset}
Let $Z$ be a minimum zero forcing set of a graph $G=(V,E)$.  Then, for any $S \subset Z$, there does not exist a set $R\subset V$ with $|R| < |S|$ and $cl(S) \subset cl(R)$.
\end{lemma}
\begin{Proof}
 Suppose for contradiction that there exists a set $R\subset V$ with $|R| < |S|$ and $cl(S) \subset cl(R)$.  Then, since $S\subset cl(S) \subset cl(R)$, it follows that the vertices in $R$ can force all the vertices in $S$ after some number of timesteps.  Since $Z$ is a zero forcing set and $R$ is capable of forcing all the vertices in $S$, it follows that $(Z \backslash S) \cup R$ is a zero forcing set with cardinality smaller than $Z$; this is a contradiction.
\hfill$\Box$
\end{Proof}

Given a graph $G=(V,E)$, a \emph{closure pair} of $G$ is an ordered pair $(S,r)$ where $S$ is the closure of some subset of $V$, and $r$ is the cardinality of a subset of $V$ whose closure is $S$. We will show that each element of the set $\mathcal{C}$ in the Wavefront algorithm is a closure pair. Note that a set whose closure is $S$ and whose cardinality is $r$ is not explicitly identified or stored in the algorithm; indeed there could be many sets with the same closure and the same cardinality.

\begin{lemma}
\label{closure_pair_lemma}
Let $G=(V,E)$ be a graph. At each step of the Wavefront algorithm applied to $G$, each element of the set $\mathcal{C}$ is a closure pair of $G$.
\end{lemma}
\begin{Proof}
We will prove the claim by induction on $R$. In line 1, $\mathcal{C}$ is initialized as a set containing a closure pair. Suppose that for some $R\geq 0$, all elements of $\mathcal{C}$ are closure pairs, and consider the next iteration of the loop on line 2 which increments $R$. 
$\mathcal{C}$ is updated only in line 8, when an ordered pair $(S',r')$ is added to $\mathcal{C}$. 
Moreover, in order for $(S',r')$ to be added to $\mathcal{C}$, the if-statement in line 7 has to be satisfied, i.e., $r'\leq R$. Thus, since the value of $R$ increases in the loop on line 2, the second element of each closure pair in $\mathcal{C}$ is no more than the current value of $R$. In lines 3 and 4, the algorithm loops over the elements of $\mathcal{C}$ and $V$. Thus, in order to show that the elements of $\mathcal{C}$ are always closure pairs, it is sufficient to show that for a fixed $R\geq 1$, and for an arbitrary iteration of the loops over $\mathcal{C}$ and $V$, the resulting $(S',r')$ which is added to $\mathcal{C}$ is a closure pair. Fix an arbitrary closure pair $(S,r)\in \mathcal{C}$ and an arbitrary vertex $v\in V$; this corresponds to fixing an arbitrary iteration of the loops. $S'$ is updated only in line 5, and by definition it is the closure of the set $S\cup N[v]$. 
Likewise, $r'$ is updated only in line 6, and by definition $r'=r+|\{v\}\backslash S| +\max\{|N(v)\backslash S|-1,0\}$.

Suppose first that $N[v]\subset S$. Then, $S'=cl(S\cup N[v])=cl(S)=S$. Since $(S,r)\in \mathcal{C}$ and $r\leq R$, the if-statement in line 7 would be false, so $(S',r')$ would not be added to $\mathcal{C}$.
Now suppose that $N[v]$ is not fully contained in $S$ and that $v$ has a neighbor $x$ outside $S$.
Since $(S,r)$ is a closure pair, there exists a set $A\subset V$ such that $cl(A)=S$ and $|A|=r\leq R$. Let $A'=A\cup ((N[v]\backslash S)\backslash \{x\})$. Then, $cl(A')=cl(S\cup N[v])$, since $A$ can force $S$, after which all-but-one neighbors of $v$ will be colored, after which $x$ can be forced by $v$, after which all remaining vertices in $cl(S\cup N[v])$ can be forced by $S\cup N[v]$. Moreover, 
$|A'|=|A|+|N[v]\backslash S|-1=r+|\{v\}\backslash S|+|N(v)\backslash S|-1=r'$.
Thus, $(S',r')$ is a closure pair, regardless of whether or not it gets added to $\mathcal{C}$. 
Finally, suppose that $N[v]$ is not fully contained in $S$ and that $v$ does not have a neighbor outside $S$, i.e., that $v\notin S$ but $N(v)\subset S$. Let $A'=A\cup \{v\}$. Then, $cl(A')=cl(A\cup \{v\})=cl(S\cup N[v])$, since $A$ can force $S$, after which all vertices in $N[v]$ will be colored, after which all remaining vertices in $cl(S\cup N[v])$ can be forced by $S\cup N[v]$. Moreover, $|A'|=|A|+1 =r+|\{v\}\backslash S|=r'$. Hence, $(S',r')$ is a closure pair, regardless of whether or not it gets added to $\mathcal{C}$. 
Thus, in every step of the algorithm, $\mathcal{C}$ is a set of closure pairs. 
\hfill$\Box$
\end{Proof}

\begin{theorem}
\label{thm:WavefrontCorrectness}
Given a graph $G=(V,E)$, the Wavefront algorithm returns $Z(G)$.
\end{theorem}

\begin{Proof}
By Lemma \ref{closure_pair_lemma}, at each step of the algorithm, $\mathcal{C}$ is a set of closure pairs of $G$. Since $G$ has a finite number of closure pairs, each loop of the algorithm is over a finite set, so the algorithm terminates. Note that at any step of the algorithm, $r'$ on line 6 is at most $n$. Thus, for $R$ large enough to satisfy the if-statement on line 7, in line 5 a neighborhood can be added to the largest closure of a closure pair in $\mathcal{C}$, creating a larger closure; eventually, all vertices can be added, so the largest closure of a closure pair in $\mathcal{C}$ will be $V$. Thus, the algorithm always returns a number. Let $r^*$ be the number returned by the algorithm, and let $Z^*$ be a minimum zero forcing set of $G$.

If $|Z^*|=n$ we are done, because the number $r^*$ returned by the algorithm always satisfies $Z(G)\leq r^*\leq R\leq n$. Thus, suppose henceforth that $Z^*\neq V$, so $Z^*$ contains some vertex $v$ together with all-but-one of its neighbors. Let $S_0=N[v]\cap Z^*$. 
By Lemma~\ref{lem:smallestsubset}, $S_0$ is a minimum cardinality set of vertices whose closure contains $cl(S_0)$.
Since $S_0$ consists of a single vertex and all-but-one of its neighbors, after at most $|S_0|$ iterations of the loop on line 2, the if-statement on line 7 will be true and $(cl(S_0),|S_0|)$ will be added to $\mathcal{C}$. If $cl(S_0) = V$ then the algorithm would terminate and return $|S_0|=|Z^*|$, so assume that $cl(S_0)\neq V$.

Let $S$ be a maximum cardinality subset of $Z^*$ such that $(cl(S),|S|)$ is added to $\mathcal{C}$ at some step of the algorithm. Note that $S\neq \emptyset$ since $|S|\geq |S_0|>0$. If $cl(S)=V$ then the algorithm would terminate at the step when $(cl(S),|S|)$ is added to $\mathcal{C}$ and return $|S|=|Z^*|$, so assume that $cl(S)\neq V$.

If $V\backslash cl(S)\subset Z^*$, then $cl(S)\cap Z^*=S$, since otherwise $(Z^*\backslash cl(S))\cup S$ would be a smaller zero forcing set than $Z^*$; thus, $|Z^*| = |S| + |V \backslash cl(S)|$. Any closure pair that is built by starting from the closure pair $(cl(S),|S|)$ and repeatedly adding neighborhoods of vertices in line 5 (and storing the intermediate closure pairs in $\mathcal{C}$ in line 8) can have its second entry be at most $|S|+|V \backslash cl(S)|=|Z^*|$; this is because no vertices from $cl(S)$ can be added at any stage, since then a closure pair with a smaller second entry would have the same closure and the if-statement on line 7 would be false, not allowing the closure pair to be added to $\mathcal{C}$. Thus, a closure pair $(V,|S|+|V \backslash cl(S)|)=(V,|Z^*|)$ will be added to $\mathcal{C}$. Moreover, no closure pair $(V,q)$ with $q<|Z^*|$ will be added to $\mathcal{C}$ since that would imply there is a zero forcing set smaller than $Z^*$. Finally, no closure pair $(V,q)$ with $q>|Z^*|$ will be added to $\mathcal{C}$, because of line 7 and because the loop on line 2 increments the values of $R$. Thus, in this case the number returned by Wavefront is $r^*=|Z^*|$.

\begin{figure}[h]
\begin{center}
\includegraphics[scale=0.35]{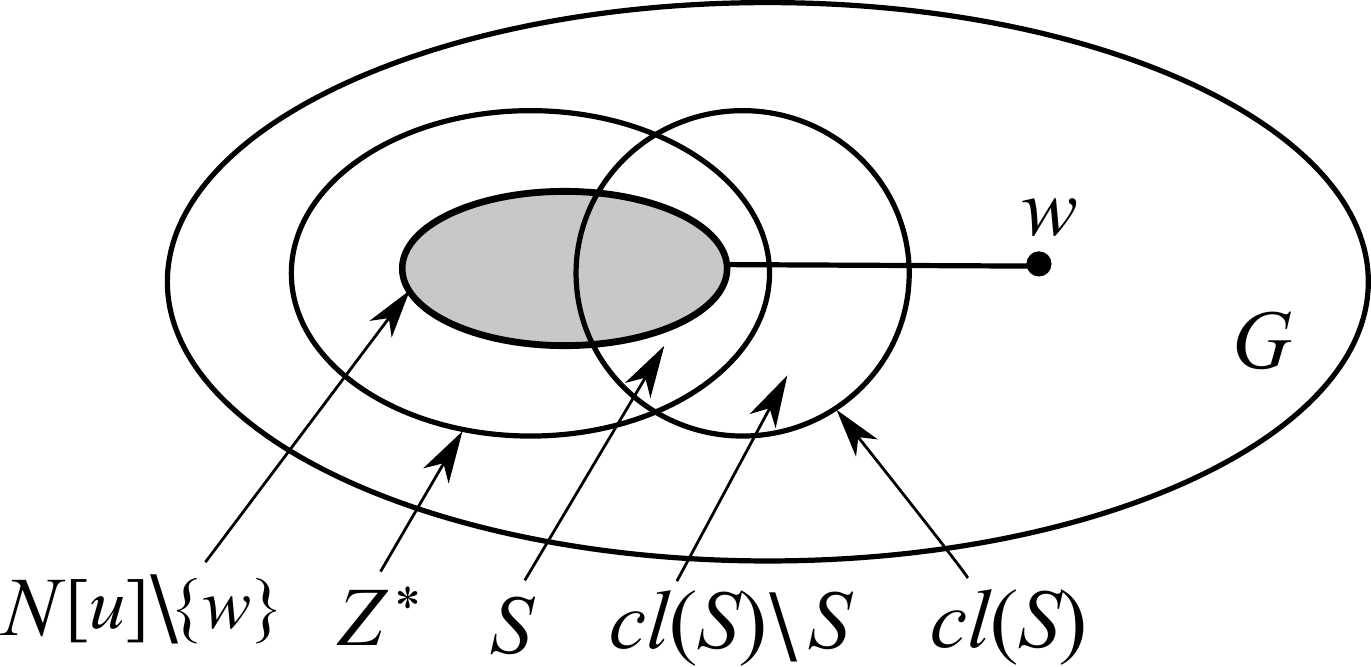}
\caption{Illustration of the proof Theorem \ref{thm:WavefrontCorrectness} in the case when $V \backslash cl(S)\not\subset Z^*$.}
\label{fig_wavefront}
\end{center}
\end{figure}

Now suppose that $V \backslash cl(S)\not\subset Z^*$; let $w$ be the first vertex outside $Z^*\cup cl(S)$ to get forced by $Z^*$, and let $u$ be the vertex which forces $w$ (given some fixed chronological list of forces). Then, since $Z^*$ is a minimum zero forcing set and $S\subset Z^*$, $Z^*$ cannot contain a vertex of $cl(S)$ which is not in $S$. Since $w$ is the first vertex forced outside of $Z^* \cup cl(S)$, it follows that $N[u]\backslash\{w\}\subset Z^*$. Thus, 
\begin{equation}
\label{eq_wavefront}
|Z^* \cap (S \cup N[u])| = |S| + |N[u] \backslash (cl(S) \cup \lbrace w \rbrace)| = |S| + |N[u] \backslash cl(S)| - 1;
\end{equation}
see Figure \ref{fig_wavefront} for an illustration. Since $w\notin cl(S)$, there must exist at least one vertex in $N[u]\backslash \{w\}$ that is not in $cl(S)$ and therefore that vertex must be in $Z^*$. Thus, 
\begin{equation}
\label{eq_wavefront2}
|Z^* \cap (S \cup N[u])|>|S|. 
\end{equation}
Note that $cl(N[u] \cup cl(S)) = cl(Z^* \cap (S \cup N[u]))$ and by (\ref{eq_wavefront}), $|S| + |N[u] \backslash cl(S)| - 1=|Z^* \cap (S \cup N[u])|$.
Thus, $(cl(N[u] \cup cl(S)), |S| + |N[u] \backslash cl(S)|-1)$ is a closure pair. Suppose this closure pair was added to $\mathcal{C}$ in some step of the algorithm.  
Then, we would have a set $\hat{S}=Z^* \cap (S \cup N[u])\subset Z^*$ that is larger than $S$, by (\ref{eq_wavefront2}), and yet $(cl(\hat{S}),|\hat{S}|)$ is added to $\mathcal{C}$. This contradicts the assumption that $S$ was the largest such set.


Thus, suppose $(cl(N[u] \cup cl(S)), |S| + |N[u] \backslash cl(S)|-1)$ is never added to $\mathcal{C}$ at any step of the algorithm.
The algorithm cannot terminate when $R<|S|+|N[u]\backslash cl(S)|-1$, since that would mean that $r^*\leq R<|S|+|N[u]\backslash cl(S)|-1\leq |Z^*|\leq |Z^*|$ (where the first inequality follows from line 7, and the last inequality follows from (\ref{eq_wavefront})), and since $(V,r^*)$ is a closure pair, there would have to exist a zero forcing set of cardinality less than $|Z^*|$, a contradiction. At the iteration $R=|S|+|N[u]\backslash cl(S)|-1$ of the loop on line 2,
since $(cl(S),|S|)$ is in $\mathcal{C}$, $S'=cl(cl(S)\cup N[u])$ will be created in line 5, and 
\begin{eqnarray*}
r'&=&|S|+|\{u\}\backslash cl(S)|+\max\{|N(u)\backslash cl(S)|-1,0\}\\
&=&|S|+|\{u\}\backslash cl(S)|+|N(u)\backslash cl(S)|-1\\
&=&|S|+|N[u]\backslash cl(S)|-1=R
\end{eqnarray*}
will be created in line 6.

If a closure pair $(cl(Q),|Q|)$ such that $cl(Q)=S'$ and $|Q|\leq R$ is already in $\mathcal{C}$ before step $R$, then $|Q|<|S| + |N[u]\backslash cl(S)| - 1$. Note that $(Z^*\backslash (cl(S \cup N[u])))\cup Q$ is a zero forcing set since $cl(Q) =cl(cl(S)\cup N[u])=cl(S \cup N[u])$, so all the vertices that are removed from $Z^*$ can be forced by $Q$. Moreover, by (\ref{eq_wavefront}), $Z^*$ contains $|S|+|N[u]\backslash cl(S)|-1$ vertices of $S\cup N[u]$, and since $S\cup N[u]\subset cl(S\cup N[u])$, $Z^*$ contains at least $|S|+|N[u]\backslash cl(S)|-1$ vertices of $cl(S \cup N[u])$. Thus, 
\begin{eqnarray*}
|(Z^*\backslash (cl(S \cup N[u])))\cup Q|&\leq& |Z^*\backslash (cl(S \cup N[u]))|+|Q|\\
&\leq& |Z^*|-(|S|+|N[u]\backslash cl(S)|-1)+|Q|<|Z^*|,
\end{eqnarray*}
a contradiction. Thus, the closure pair $(cl(Q),|Q|)$ does not exist, so the if-statement on line 7 is true and $(cl(N[u] \cup cl(S)), |S| + |N[u] \backslash cl(S)|-1)$ must is added to $\mathcal{C}$; this is a contradiction.
\hfill$\Box$
\end{Proof}

\begin{theorem}
\label{thm:WavefrontComplexity}
At any step $s$ of the Wavefront algorithm, 
\begin{equation}
\label{C_bound}
|\mathcal{C}| \leq \sum_{i=1}^s \binom{n}{i},
\end{equation}
and this bound is tight.
\end{theorem}
\begin{Proof}
Multiple sets that have the same closure are not added to $\mathcal{C}$. Since all permutations of a set have the same closure, at most ${n \choose s}$ new sets can be added to $\mathcal{C}$ in step $s$. Thus, after step $s$, $|\mathcal{C}|$ is bounded as in (\ref{C_bound}). The worst case performance of Wavefront is realized in empty graphs (after $n$ steps), since the closure of any set of vertices is the set itself, and since every combination of vertices must be checked before a zero forcing set is found.
\hfill$\Box$\end{Proof}

As shown in Theorem~\ref{thm:WavefrontComplexity}, in the worst case, the Wavefront algorithm is no better than enumerating all possible subsets of vertices; graphs in which very few vertices can be forced by sets with fewer than $Z(G)$ elements (e.g. stars) also lead to poor performance. However, Wavefront performs much better than the brute force algorithm when the closures of subsets of vertices are larger than the original subsets. This improvement comes from the fact that when some vertices being forced have no uncolored neighbors, they are no longer possible choices to add to the sets in $\mathcal{C}$. 

While Wavefront could potentially be modified to create connected forcing sets, such a modification would eliminate the computational advantages of the algorithm.  Wavefront only stores optimal forcing sets for certain subgraphs of $G$, and then builds the optimal forcing sets of larger subgraphs by adding neighborhoods of vertices containing an uncolored vertex. However, an optimal forcing set for a subgraph may not be connected to other vertices that must be added in order to force the entire graph. Thus, to be useful for finding connected forcing sets, Wavefront would have to store more than just the optimal forcing set for each subgraph, and its performance would suffer as a result. For these reasons, Wavefront will not be a viable method for solving the connected zero forcing problem without significant alterations.

\subsection{Branch-and-Bound Algorithm}
\label{sec:branch and bound}

We now present a combinatorial branch-and-bound algorithm for connected zero forcing; this algorithm is based on a variant of reverse search. It was shown by Avis and Fukuda \cite{AVIS_FUKUDA} that the reverse search technique generates all connected induced subgraphs of a graph; we include a proof below for completeness.

\begin{algorithm2e}
\KwData{Connected graph $G=(V,E)$}
\KwResult{Connected forcing number of $G$}
\SetKwProg{Fn}{function}{}{end}
$R \leftarrow V$, $S \leftarrow \emptyset$, $\ell \leftarrow |V|$, $Z_c \leftarrow |V|$\;
\Fn{\emph{ConnectedForcingSet}$(R,S,\ell)$}{
$C\leftarrow R \cap N(S)$\;
\textbf{if} $S=\emptyset$ \textbf{then} $C\leftarrow R$\;
\If{$C=\emptyset$ \textbf{\emph{and}} $cl(S)=V$}{
    $\ell\leftarrow |S|$\;

\textbf{if} $|S|<Z_c$ \textbf{then} $Z_c\leftarrow |S|$\;
  }

\Else{
  Choose any $v \in C$\;
  ConnectedForcingSet($R \backslash\{v\}$, $S$, $\ell$)\;

\textbf{if} $|S| < \ell-1$ \textbf{then} ConnectedForcingSet($R \backslash\{v\}$, $S \cup \{v\}$, $\ell$)\;
 
 }
 }
\Return $Z_c$\;
\caption{Branch-and-bound algorithm for Connected Zero Forcing \label{alg:BandBZeroForcing}}
\end{algorithm2e}


\begin{theorem}
Given a graph $G=(V,E)$, Algorithm \ref{alg:BandBZeroForcing} returns $Z_c(G)$.
\end{theorem}
\begin{Proof}
In Algorithm \ref{alg:BandBZeroForcing} and throughout this proof, $S$ denotes the vertex set of a partially constructed subgraph, $R$ denotes the set of vertices not yet considered, and $C$ denotes the set of candidate vertices that could be added to $S$. Algorithm \ref{alg:BandBZeroForcing} is based on the following Subgraph function, which enumerates all subgraphs of $G$. In particular, this function recursively adds or does not add to $S$ a vertex $v\in R$ which has not yet been considered. When initialized with $R=V$ and $S=\emptyset$, this process defines a binary tree $T$, where the choice of whether or not $v$ gets added to $S$ gives two branches of a subtree of $T$ descending from the node representing $S$. Then, the leaves of $T$ are the subgraphs of $G$, and are visited depth-first by the recursion.

\setlength\algotitleheightrule{0pt}
\setlength\algoheightrule{0pt}
\begin{algorithm2e}[h!]
\SetKwProg{Fn}{function}{}{end}
\Fn{\emph{Subgraph}$(R,S)$}{
\textbf{if} $R=\emptyset$ \textbf{then} \textbf{print} $S$\;
\Else{
Choose any $v \in R$\;
Subgraph($R \backslash\{v\}$, $S$)\;
Subgraph($R \backslash\{v\}$, $S \cup \{v\}$)\;
 }
 }
\end{algorithm2e}

To generate the connected induced subgraphs of $G$, the Subgraph function can be modified into the ConnetedSubgraph function as follows: instead of choosing any vertex $v\in R$ to branch on, the vertex $v$ is chosen to be in $C=R\cap N(S)$ (and in the first level of the search tree, when $S=\emptyset$, the choice of $v$ is unconstrained). This assures that at each step, $S$ is connected, and the leaves of the search tree $T$ are the connected induced subgraphs of $G$. If the set of candidate vertices $C$ is empty, there are no other connected induced subgraphs in the subtree descending from $S$, so that subtree is pruned. 

\begin{algorithm2e}[h!]
\SetKwProg{Fn}{function}{}{end}
\Fn{\emph{ConnectedSubgraph}$(R,S)$}{
$C\leftarrow R \cap N(S)$\;
\textbf{if} $S=\emptyset$ \textbf{then} $C\leftarrow R$\;
\textbf{if} $C=\emptyset$ \textbf{then print} $S$\;
\Else{
  Choose any $v \in C$\;
ConnectedSubgraph($R \backslash\{v\}$, $S$)\;
ConnectedSubgraph($R \backslash\{v\}$, $S \cup \{v\}$)\;
 }
 }
\end{algorithm2e}

Finally, to visit the minimum connected forcing sets of $G$, in  the ConnectedForcingSet function of Algorithm \ref{alg:BandBZeroForcing}, each subtree of $T$ only includes subsets of vertices that are larger than the subset represented by the root of the subtree. Thus, once a connected zero forcing set of size $\ell$ is found, all subsequent subtrees that lead to subsets of size at least $\ell$ can be pruned. Hence, the correctness of Algorithm \ref{alg:BandBZeroForcing} follows from the fact that all connected induced subgraphs whose order is less than or equal to the cardinality of an already-discovered connected zero forcing set are enumerated. 
\hfill$\Box$
\end{Proof}

Since a graph $G$ may have exponentially-many connected induced subgraphs with fewer than $Z_c(G)$ vertices, in the worst case this algorithm is no better than brute force. However, as with Wavefront, Algorithm \ref{alg:BandBZeroForcing} performs much better in practice, when $G$ has relatively few connected induced subgraphs or when large parts of the search tree are pruned.

\section{Integer Programming Approaches}
\label{sec:IPMethods}
In this section, we describe several integer programming formulations and solution strategies for computing the zero forcing and connected forcing numbers of a graph, and for finding the smallest set which forces a graph within a specified number of timesteps. The presented formulations come from two distinct perspectives on zero forcing. The first perspective is a straightforward model of zero forcing as a dynamic graph infection process. This approach incorporates the dynamic nature of the forcing process by using the vertices forced at each timestep to determine the vertices that can be forced in the next timestep. The second perspective uses the theory of zero forcing forts introduced by Fast and Hicks \cite{ForcingTheoryPaper}, and models zero forcing as a type of set-covering problem which does not depend on timesteps.

\subsection{Infection Perspective}
In the formulation of zero forcing as a dynamic process, each edge of the given graph $G=(V,E)$ is replaced by two directed edges with opposite directions. A binary variable $s_v$ indicates whether vertex $v$ is in the forcing set; an integer variable $x_v$ ranging in $\{0,\ldots,T\}$ indicates at which timestep vertex $v$ is forced, where $T$ is the maximum difference between the forcing times of two vertices; finally, a binary variable $y_e$ for each directed edge $e=(u,v)$ indicates whether $u$ forces $v$.  The notation $\delta^-(v)$ refers to the set of edges pointing towards node $v$.

\begin{Model_float}
\begin{Model}{IP model for Zero Forcing based on infection}
\label{Def:InfIP}
\begin{eqnarray}
\nonumber \min && \sum_{\mathclap{v \in V}}s_{v}\\
\emph{s.t.:}&& s_v + \sum_{\mathclap{e\in \delta^- (v)}}y_e = 1 \qquad \forall v \in V \label{M1C1}\\ 
&& x_u - x_v + (T+1)y_e \leq T\qquad \forall e=(u,v) \in E \label{M1C2}\\ 
&& x_w - x_v + (T+1)y_e \leq T \qquad \forall e=(u,v) \in E, \forall w \in N(u)\backslash\{v\}\label{M1C3}\\ 
\nonumber&& x\in\{0,\ldots,T\}^n, s\in \{0,1\}^n, y\in\{0,1\}^m
\end{eqnarray}
\end{Model}
\end{Model_float}

\begin{theorem}
 The optimum of Model~\ref{Def:InfIP} is equal to $Z(G)$.
\end{theorem}
\begin{Proof}
Let $Z$ be a zero forcing set, fix some chronological list of forces, and let $\mathcal{F}$ be the associated set of forcing chains. Since $Z$ is a zero forcing set, each vertex $v$ of $G$ is either in $Z$ (i.e. $s_v = 1$) or is forced by some other vertex of $G$ (i.e. $y_e = 1$ and $v$ is the head of $e$).  Thus, constraint (\ref{M1C1}) must be satisfied.  Now, let $x_v$ be the timestep in which $v$ is forced. Since a vertex cannot force until all-but-one of its neighbors are forced, it follows that for every edge $e=(v,w)$ for which $y_e =1$, $v$ must be forced before $w$ and thus $x_v < x_w$.  Likewise, $x_i < x_w$ for all neighbors $i$ of $v$. Thus, constraints (\ref{M1C2}) and (\ref{M1C3}) are satisfied.  If $y_e=0$, then constraints (\ref{M1C2}) and (\ref{M1C3}) are satisfied since $T$ is the maximum difference between the forcing times of two vertices.  Thus, the constraints are valid for any zero forcing set and associated set of forcing chains. 

Conversely, let $(s,x,y)$ be a feasible solution of Model~\ref{Def:InfIP} for the given graph $G$, and let $Z$ be the set of all vertices for which $s_v = 1$. By constraint (\ref{M1C1}), each vertex $v$ is either in $Z$, or is the head of exactly one edge $e=(u,v)$ with $y_e=1$. Consider an edge $e=(u,v)$ for which $y_e=1$; by  constraints (\ref{M1C2}) and (\ref{M1C3}), there must be some integer $x_v \in \lbrace0,\ldots,T\rbrace$ such that $x_i+1 \leq x_v$ for $i \in N[u] \backslash \lbrace v \rbrace$. By interpreting the $x_i$ variable as the timestep in which the vertex $i$ is forced for each vertex with $s_i = 0$, it follows that there must exist some timestep $x_v\in \lbrace0,\ldots,T\rbrace$ such that if $v$ is forced by $u$ in timestep $x_v$, then $u$ and all its neighbors except $v$ have been forced in previous timesteps. Thus, each vertex is the tail of at most one edge $e$ with $y_e=1$, and so the edges for which $y_e=1$ define a set $\mathcal{F}$ of directed paths which have one end-vertex in $Z$. Since every vertex of $G$ is either in $Z$ or is the head of an edge with $y_e=1$ and is therefore part of a path in $\mathcal{F}$, it follows that $Z$ is a zero forcing set of $G$ and $\mathcal{F}$ is a set of forcing chains associated with $Z$. 
\hfill$\Box$\end{Proof}

Note that for any graph, since at least one vertex is forced at each timestep, it follows that $T<n$; it is possible to further bound $T$ if the graph is assumed to have certain properties (see, e.g., \cite{throttling,iteration_index,ForcingTheoryPaper,proptime1,proptime2} for results on the propagation time $T$).

The main advantage of Model~\ref{Def:InfIP} is that the numbers of constraints and variables in this formulation are polynomial in $n$; therefore, the integer program can be solved directly, without delayed row or column generation. Another useful feature of this model is that it not only finds the zero forcing number and a minimum zero forcing set of $G$, but it also gives a set of forcing chains associated with the forcing set. However, the downfall of Model~\ref{Def:InfIP} is its reliance on constraints (\ref{M1C2}) and (\ref{M1C3}), which are of big-$M$ form; these constraints lead to poor performance for this model (except on very sparse graphs, as shown in Section~\ref{sec:Zero_Forcing_Results}).

Model~\ref{Def:InfIP} is easily adapted for a different purpose -- namely, for finding a zero forcing set which forces $G$ within a specified number of timesteps, and has minimum cardinality subject to that property. This result can be achieved by fixing $T$ in Model~\ref{Def:InfIP} to be the maximum acceptable number of timesteps to force the graph. As mentioned above, the zero forcing propagation time has been previously investigated (cf. \cite{throttling,iteration_index,ForcingTheoryPaper,proptime1,proptime2}) from a combinatorial standpoint, but to our knowledge, this is the first computational tool for this problem for general graphs. Our experiments show that smaller $T$ values cause Model~\ref{Def:InfIP} to solve faster and handle larger graphs. 

\subsection{Fort Covering Perspective}
\label{sec:Fort_Covering_Perspective}
Our next formulation models zero forcing as a set-covering problem which does not depend on timesteps and does not rely on big-$M$ constraints. In Model~\ref{Def:FortIP}, the binary variable $s_v$ again indicates whether vertex $v$ is in the zero forcing set; $\mathcal{B}$ is the set of all forts in the given graph $G=(V,E)$.

\begin{Model_float}
\begin{Model}{IP model for Zero Forcing based on forts}
\label{Def:FortIP}
\begin{eqnarray}
\nonumber    \min && \sum_{\mathclap{v \in V}}s_{v}\\
    \emph{s.t.:} && \sum_{\mathclap{v \in B}}s_v \geq 1  \qquad \forall B \in \mathcal{B}\label{M2C1}\\
\nonumber    &&s\in\{0,1\}^n
\end{eqnarray}
\end{Model}
\end{Model_float}

\begin{theorem}
\label{thm4}
 The optimum of Model~\ref{Def:FortIP} is equal to $Z(G)$.
\end{theorem}
\begin{Proof}
Let $s$ be a feasible solution of Model~\ref{Def:FortIP}, and let $Z$ be the set of all vertices $v$ for which $s_v=1$. Suppose for contradiction that $Z$ is not a zero forcing set of $G$, which means $cl(Z) \neq V$. If any vertex $u\in cl(Z)$ was adjacent to exactly one vertex $v\in V\backslash cl(Z)$, then $u$ could force $v$, contradicting the definition of $cl(Z)$. Thus, $V\backslash cl(Z)$ is a fort, and it does not contain any vertex of $Z$; this means constraint (\ref{M2C1}) is violated.  It follows that $Z$ must be a zero forcing set of $G$.

Conversely, let $Z$ be a zero forcing set of $G$. Suppose for contradiction that there exists a fort $F$ which does not contain any element of $Z$. In order for the first vertex $v$ of $F$ to be forced, at some timestep, $v$ must be the only neighbor of some colored vertex outside $F$. However, since $F$ is a fort, any vertex outside $F$ which is adjacent to $v$ is also adjacent to another (uncolored) vertex in $F$. Thus, $v$ cannot be forced, which contradicts $Z$ being a forcing set. It follows that every fort contains an element of $Z$, so $Z$ is a feasible solution of Model~\ref{Def:FortIP}. \hfill$\Box$\end{Proof}

In contrast to Model~\ref{Def:InfIP}, Model~\ref{Def:FortIP} has the advantage of not having big-$M$  constraints. However, the main issue with Model~\ref{Def:FortIP} is that since a graph could have an exponential number of forts (in the sense that there exist families of graphs with $n$ vertices and $\Omega(2^n)$ forts, e.g., $K_n$), the solution methodologies for Model~\ref{Def:FortIP} must use constraint generation (see \cite{DantFulkJohn} or \cite{NemhauserWolsey} for an introduction to this approach). In constraint generation, a relaxed master problem (RMP) is obtained by omitting a set of constraints (in this case, the omitted constraints are the fort cover constraints (\ref{M2C1})); then, the RMP is solved, a set of violated constraints from the full model is added to the RMP, and this process is repeated until there are no more violated constraints.  

The usefulness of constraint generation depends on the development of a practical method for finding violated constraints. A few high-quality constraints can subsume a large number of low-quality constraints, but the former are usually expensive to find. Thus, there is a trade-off between the time spent on finding constraints, and the number of constraints that have to be found. In the remainder of this section, we consider several different methods for finding violated constraints. 

A quick and naive way to generate constraints is the following: if $S$ is a solution to a RMP, the vertex complement of the closure of $S$, i.e. $V \backslash cl(S)$, is a violated fort cover constraint. We will refer to this method of generating constraints as the \emph{closure complement method}.

Another method to find violated forts is to use the auxiliary integer program given in Model~\ref{Def:FindFortIP}. For this model, we define the set $S$ to be the set of all vertices for which $s_v = 1$ in the current optimal solution of a RMP for Model~\ref{Def:FortIP}. Note that since the value for each $s_v$ is taken from the current optimal solution of the RMP, $S$ is 
constant for Model~\ref{Def:FindFortIP}. The $x_v$ variables indicate whether vertex $v$ is in the fort.

\begin{Model_float}
\begin{Model}{IP model for finding forts}
\label{Def:FindFortIP}
\begin{eqnarray}
\nonumber\min&& \sum_{\mathclap{v \in V}}x_{v}\\
\emph{s.t.:}&& \sum_{\mathclap{v \in V}}x_v \geq 1\label{M3C1}\\
&&x_w - x_v +\sum_{\mathclap{a \in N(w)\backslash \{v\}}} x_a \geq 0\qquad\forall(v,w) \emph{ with } v \in V, w \in N(v)\label{M3C2}\\
&&x_v = 0\qquad\forall v \in cl(S)\label{M3C3}\\
\nonumber&& x\in\{0,1\}^n
\end{eqnarray}
\end{Model}
\end{Model_float}

\begin{theorem}
Let $S$ be a feasible solution of a RMP of Model \ref{Def:FortIP}.  Model~\ref{Def:FindFortIP} finds a minimum size violated fort with respect to $S$.
\end{theorem}
\begin{Proof}
 Let $B$ be a violated fort of $G$; let $x_v=1$ for $v \in B$, and $x_v=0$ for $v\notin B$. Since a fort is non-empty by definition, $B$ must contain at least one vertex; therefore, constraint (\ref{M3C1}) of the model is satisfied. Again by the definition of a fort, any neighbor of a vertex in $B$ must either be in $B$ or have at least one other neighbor in $B$; therefore, constraint (\ref{M3C2}) is satisfied. Since $B$ is a violated fort, no vertex in $B$ can be in $cl(S)$; therefore, constraint (\ref{M3C3}) is satisfied.
 
Conversely, let $x$ be a solution of Model~\ref{Def:FindFortIP}, and let $B$ be the set of vertices of $G$ for which $x_v=1$. By constraint (\ref{M3C1}), $B$ is not empty.  By constraint (\ref{M3C2}), every neighbor of a vertex in $B$ must either be in $B$, or have at least two  neighbors in $B$ (note that if $v$ is in $B$, then $v$ is one of the neighbors of $w$ which is in $B$).  Thus, $B$ is a fort of $G$. Furthermore, by constraint (\ref{M3C3}), no vertex of $B$ is in $cl(S)$; therefore $B$ is a violated fort with respect to the solution $S$. 
\hfill$\Box$\end{Proof}

Model~\ref{Def:FindFortIP} separates violated constraints for Model~\ref{Def:FortIP}. There is precedent in literature for using an integer programming separation method (see, e.g., \cite{AVELLA_BOCCIA_VASILYEV,FISCHETTI_LODI}).  In our computational experiments, Model~\ref{Def:FindFortIP} solved relatively quickly, and the forts found using this method are smaller and therefore more effective at solving Model~\ref{Def:FortIP} than those found by the closure complement method. 

Instead of adding minimum forts using the IP in Model \ref{Def:FindFortIP}, we can instead add minimal forts, in polynomial time, by finding a maximal non-forcing set $M$ that contains a solution $S$ of the RMP; then,  $V(G)\backslash M$ is a minimal fort that can be added as a constraint. To find a maximal non-forcing set $M$ that contains $S$, a greedy approach can be used whereby $M$ is initialized as $cl(S)$ and the vertices in $V(G)\backslash M$ are searched for a vertex that can be added to $M$ without forming a forcing set; if such a vertex is found, it is added to $M$ and the search repeats until no more vertices can be added. We will call this method of generating minimal forts the \emph{maximal closure method}.  In some cases, the minimal forts 
will also be minimum, and we will get the effectiveness of the minimum forts without the cost of solving an IP.  However, as graphs become larger and 
more complex, it becomes more likely that the minimal forts will not be minimum.  These observations are borne out in our computational results, which show 
that the maximal closure method performs well on small graphs, but is beaten for larger graphs (in terms of number of instances that can be solved to optimality) by the models that find minimum forts.

Since Model~\ref{Def:FortIP} is a set-covering problem, we can use the theory of Balas and Ng~\cite{BALAS_NG} on the set-covering polytope to explain why the forts generated by Model~\ref{Def:FindFortIP} are more effective than those found by the closure complement method. In particular, we restate a theorem from \cite{BALAS_NG} in terms of forts; this result gives necessary and sufficient conditions for inequalities with a right-hand-side of one to be facet inducing.

\begin{theorem}\emph{\cite{BALAS_NG}}
\label{thm:FortFacets}
 Given a fort $B$, the inequality $\sum_{v \in B}s_v \geq 1$ defines a facet of the zero forcing polytope if and only if the following two conditions hold:
\begin{eqnarray}
&&  \nexists A\subset B \text{ s.t. } A \text{ is a fort,}\label{condition1}\\
&&  \forall v \notin B, \exists w \in B \text{ s.t. if } A \subset B \cup \{v\}\text{ is a fort and }v \in A, \text{ then } w\in A\label{condition2}.
\end{eqnarray}
%

\end{theorem}

Note that in condition (\ref{condition2}), the choice of $w$ depends on $v$ and must remain the same for all forts $A \subset B \cup \{v\}$.  Condition (\ref{condition1}) explains why Model~\ref{Def:FindFortIP} performs better than the closure complement method. The latter makes no effort to minimize the size of the generated forts; thus, they are unlikely to satisfy (\ref{condition1}) and be facet inducing.  On the other hand, Model~\ref{Def:FindFortIP} finds minimum size violated forts, which satisfy (\ref{condition1}). However, condition (\ref{condition2}) is not necessarily satisfied by either method.

This motivates further investigation of ways to add facet-inducing inequalities to Model~\ref{Def:FortIP}, which we address next. Note that if (\ref{condition2}) is violated for a fort $F$, then there exist $p$ forts $A_1,\ldots,A_p \subset F \cup \{v\}$ such that there is a vertex $v$ with $v \in A_i$, $1\leq i\leq p$, but $\bigcap_{i \in [p]} (A_i \backslash \{v\}) = \emptyset$.  Observe also that the fort constraints given by $F$ and $A_i$ for $1\leq i \leq p$ can be combined to give the valid cut $\sum_{i \in F \cup \{v\}}x_i \geq 2$. This valid cut is found by Chv\'{a}tal-Gomory rounding: we first sum the fort constraints corresponding to $F$ and all the $A_i$; since $\bigcap_{i \in [p]} A_i = \emptyset$, the coefficient of each vertex variable in the sum is at most $p$, but the right-hand-side is $p+1$.  Thus, the valid cut can be obtained by dividing through by $p$ and taking the ceiling of each coefficient. Moreover, the magnitude of $p$ is bounded as follows.  
\begin{theorem}
 \label{thm:NumFortsinIP}
Suppose there exist $p$ forts $A_1,\ldots,A_p \subset F \cup \{v\}$ such that there is a vertex $v$ with $v \in A_i$ for $1\leq i \leq p$, but $\bigcap_{i \in [p]} (A_i \backslash \{v\}) = \emptyset$. Then $p$ can be chosen to be at most $|F|$.
\end{theorem}
\begin{Proof}
Suppose there exist $q > |F|$ forts that satisfy the required properties. Then, for each vertex $w \in F$, choose one fort among $A_1,\ldots,A_q$ that does not contain $w$; since $\bigcap_{i \in [q]} (A_i \backslash \{v\}) = \emptyset$, such a fort must exist for each $w$. This collection consists of at most $|F|$ forts, whose intersection is empty except for $v$.  Thus, it is possible to choose $p\leq |F|$ forts which satisfy the required properties.
\hfill$\Box$\end{Proof}

Given Theorem \ref{thm:NumFortsinIP}, Model~\ref{Def:FacetIP} can be used to check whether a fort generated by Model~\ref{Def:FindFortIP} is facet inducing.  If the generated fort is not facet inducing, then the valid cut generated as described above can be added instead of a fort constraint. In Model~\ref{Def:FacetIP}, the variable $z_{ij}$ indicates whether vertex $j$ is chosen to be in fort $i$, and the variable $y_i$ indicates whether fort $i$ is empty.

\begin{Model_float}
\begin{Model}{IP model for checking if a fort is facet inducing}
\label{Def:FacetIP}
\begin{eqnarray}
\nonumber\min &&\sum_{\mathclap{i \in [|F|]}}y_i\\
\emph{s.t.:}&& \sum_{\mathclap{v \in V\backslash F}}x_{v} = 1\label{M4C1}\\
&& \sum_{\mathclap{v\in V \backslash F}}z_{iv} = y_i \qquad\qquad \forall i \in [|F|]\label{M4C2}\\
&& z_{iv} \leq x_v  \qquad\qquad \forall i \in [|F|], \forall v \in V\backslash F\label{M4C3}\\
&& \sum_{\mathclap{i \in [|F|]}}z_{iw} \leq \sum_{\mathclap{i \in [|F|]}}y_i -1 \qquad\qquad \forall w \in F\label{M4C4}\\
&& z_{iw} -z_{iu} +\sum_{\mathclap{a \in N(w)\backslash \{u\}}} z_{ia} \geq 0 \quad u \in V, w \in N(v), \forall i \in [|F|]\label{M4C5}\\
&& z_{iw} \leq y_i \qquad\qquad \forall i \in [|F|], \forall w \in V\label{M4C6}\\
\nonumber &&z\in \{0,1\}^{n|F|}, y \in \{0,1\}^{|F|}
\end{eqnarray}
\end{Model}
\end{Model_float}

\begin{theorem}
 If Model~\ref{Def:FacetIP} is infeasible, and $F$ is a minimum size fort, then the fort $F$ is facet inducing.  If Model~\ref{Def:FacetIP} has an optimal solution, then the set of forts with $y_i = 1$ shows that $F$ is not facet inducing by condition (\ref{condition2}) of Theorem~\ref{thm:FortFacets}.
\end{theorem}
\begin{Proof}
Suppose a minimum size fort $F$ is not facet inducing. Since $F$ has minimum size, it must satisfy condition (\ref{condition1}), so $F$ can only violate (\ref{condition2}). Therefore, there must exist $p$ forts $A_1,\ldots,A_p \subset F \cup \{v\}$ such that there is a vertex $v$ with $v \in A_i$ for $1\leq i\leq p$ but $\bigcap_{i \in [p]} (A_i \backslash \{v\}) = \emptyset$. By Theorem~\ref{thm:NumFortsinIP}, we can assume that $p \leq |F|$.  Let $y_i = 1$ for $1\leq i \leq p$, and let $y_i=0$ otherwise; let $z_{iw} = 1$ if $w$ is in fort $A_i$ for $1\leq i\leq p$ and $z_{iw} = 0$ otherwise; let $x_v=1$ and $z_{iv}=1$ if $1\leq i\leq p$.  All other variables are set to $0$.  Now, observe that (\ref{M4C1}) is satisfied because $x_w = 0$ for all $w \neq v$, and $x_v = 1$. Constraint (\ref{M4C2}) is satisfied because each fort $A_i$ contained $v$, and (\ref{M4C3}) is satisfied because $z_{iv}$ is either $0$ or $1$ and $x_v = 1$. Constraint (\ref{M4C4}) is satisfied because $\bigcap_{i \in [p]} (A_i \backslash \{v\}) = \emptyset$. Constraint (\ref{M4C5}) is satisfied because each $A_i$ was a fort, and (\ref{M4C6}) is satisfied because the $z$ variables are chosen to be 1 only for the forts with $y$ variables chosen to be 1.  Hence, $(z,y)$ is a feasible solution to Model~\ref{Def:FacetIP}. Thus, if Model~\ref{Def:FacetIP} is infeasible, then the fort $F$ must be facet inducing.
 
Conversely, if Model~\ref{Def:FacetIP} has an optimal solution, then defining $A_i = \{w \in V \colon z_{iw} = 1 \}$ gives a set of forts which shows that $F$ does not satisfy condition (\ref{condition2}) of Theorem~\ref{thm:FortFacets}, and hence $F$ is not facet inducing.
\hfill$\Box$\end{Proof}

Instead of guaranteeing that a fort constraint is facet inducing using Model~\ref{Def:FacetIP}, we can also determine this characteristic in a heuristic manner. In particular, Model~\ref{Def:FacetIP} can be simplified by limiting the number of forts that can be chosen, i.e., requiring at most 2 forts instead of $|F|$ forts.  Our preliminary testing showed that this simplification generally has better performance than the full model.  

%

\subsection{Fort Covering Extended}

Another way to improve the formulation of Model~\ref{Def:FortIP} comes from the observation that any zero forcing set must contain some vertex together with all-but-one of its neighbors. This idea is expressed in Model~\ref{Def:Fort_NeighborhoodIP}, with the addition of binary variables $z_v$, which indicate that vertex $v$ and all-but-one of its neighbors belong to a zero forcing set. Hence, the $z_v$ variables have a cost of $|N(v)|$ in the objective function, and at least one of them is required to be positive (enforced by constraint (\ref{M6C2})). Note that if a vertex $w$ is in the closure of $N[v]$ and $z_v$ is positive, then $w$ will be forced in the corresponding solution. Therefore, $s_w$ and $z_v$ will never both be positive; this is enforced by constraint (\ref{M6C3}). Finally, the fort constraints (constraint (\ref{M2C1}) of Model~\ref{Def:FortIP}) are modified to (\ref{M6C1}) to allow satisfaction by $z_v$ variables. Despite the increased number of variables, Model~\ref{Def:Fort_NeighborhoodIP} performs better in our experiments than Model~\ref{Def:FortIP}.  

\begin{Model_float}
\begin{Model}{IP model for Zero Forcing based on forts with neighborhood variables}
\label{Def:Fort_NeighborhoodIP}
\begin{eqnarray}
\nonumber\min && \sum_{\mathclap{v \in V}}|N(v)|z_{v} + \sum_{\mathclap{v \in V}}s_{v}\\
\emph{s.t.:}&& \sum_{\mathclap{v \in B}}\big(s_v + \sum_{\mathclap{v \in cl(N[w])}}z_w\big) \geq 1\qquad \forall B \in \mathcal{B}\label{M6C1}\\
&&\sum_{\mathclap{v \in V}}z_v \geq 1\label{M6C2}\\
&&s_w + z_v \leq 1\qquad \forall v \in V, \forall w \in cl(N[v])\label{M6C3}\\
\nonumber&&s\in \{0,1\}^n,z \in \{0,1\}^n
\end{eqnarray}
\end{Model}
\end{Model_float}

Given the additional variables in Model~\ref{Def:Fort_NeighborhoodIP}, the constraint generation models must also be expanded to generate violated forts. Instead of minimizing the number of vertices in the fort as in Model~\ref{Def:FindFortIP}, our experiments showed better performance when we minimize the number of vertices in the fort that are adjacent to vertices outside of the fort. Such \emph{minimum border forts} can be found using the integer program in Model~\ref{Def:FindMinimalBorderFortIP}. In this model, $b_v$ is a binary variable that indicates whether the vertex $v$ is adjacent to vertices outside of the fort. $S$ is the set containing every vertex $v$ such that either $s_v=1$ in the current solution, or $v$ is in $N[w]$ for some $w$ with $z_w = 1$ in the current solution. Constraint (\ref{M7C3}) ensures that the $b_v$ variables correctly indicate whether $v$ is on the border of the fort.

\begin{Model_float}
\begin{Model}{IP model for finding minimum border forts}
\label{Def:FindMinimalBorderFortIP}
\begin{eqnarray}
\nonumber \min && \sum_{\mathclap{v \in V}}b_{v}\\
\emph{s.t.:}&& \sum_{\mathclap{v \in V}}x_v \geq 1\label{M7C1}\\
&&x_w -x_v +\sum_{\mathclap{a \in N(w)\backslash \{v\}}} x_a \geq 0\qquad \forall v,w \emph{ with } v \in V, w \in N(v)\label{M7C2}\\
&&|N(v)|x_v -|N(v)|b_v - \sum_{\mathclap{a \in N(v)}}x_a \leq 0\qquad \forall v \in V\label{M7C3}\\
&&x_v = 0\qquad \forall v \in cl(S)\label{M7C4}\\
\nonumber &&x\in \{0,1\}^n, b\in \{0,1\}^n
\end{eqnarray}
\end{Model}
\end{Model_float}

%

\subsection{Fort Covering for Connected Zero Forcing}
In this section, we adapt the models introduced previously to the connected forcing problem, by adding constraints to enforce connectivity on the chosen zero forcing set. We focus on adding connectivity constraints to Model~\ref{Def:FortIP}, because it is the best performing model from the previous sections that allows us to ensure connectivity.  The neighborhood variables in Model~\ref{Def:Fort_NeighborhoodIP} make it difficult to enforce connectivity since one vertex in the neighborhood is not in the zero forcing set. 

Drawing from the literature on connected dominating sets and connected power dominating sets, there are multiple ways of enforcing connectivity in integer programs. Fan and Watson \cite{FanWatson} compared Miller-Tucker-Zemlin (MTZ) constraints, Martin constraints, single-commodity flow constraints, and multi-commodity flow constraints, and found that the MTZ constraints provide the best computational performance for both the connected dominating set and connected power dominating set problems. Another method of enforcing connectivity is to add $a$,$b$-separation cutting planes when needed, in order to cut off disconnected solutions.  This method has been used for connected dominating sets \cite{BuchananSangSungButenkoPasiliao}, Steiner trees \cite{FischettiConnectivity}, and forest planning problems \cite{CarvajalConnectivity} (see also \cite{WangBuchananButenko} for conditions that cause such inequalities to induce facets of the connected subgraph polytope). In view of these results, we explore the effectiveness of MTZ constraints and $a$,$b$-separation inequalities for enforcing connectivity in the connected zero forcing problem.

MTZ constraints were originally introduced by Miller, Tucker, and Zemlin \cite{MillerTuckerZemlin} in relation to the Traveling Salesman Problem.  The basic idea of MTZ constraints is to enforce the existence of a directed spanning tree in the subgraph induced by the chosen vertices.  Our implementation follows Fan and Watson's \cite{FanWatson} explanation of the method introduced by Quint\={a}o, da Cunha, Mateus, and 
Lucena \cite{QuintaoDaCunhaMateusLucena}. Two new vertices labeled $\alpha$ and $\beta$ are added to the given graph $G=(V,E)$, along with a set $E_{new}$ of edges containing a directed edge from each of the two new vertices to all the original vertices; $E_{new}$ also contains a directed edge from $\alpha$ to $\beta$. In the modified graph, the vertices which are not chosen to be in a forcing set will have a positive edge variable coming into them from $\alpha$, while $\beta$ will have a positive edge variable going to the root of the directed spanning tree of the chosen connected zero forcing set.

Model \ref{Def:MTZIP} combines Model \ref{Def:FortIP} with the MTZ constraints.  In Model \ref{Def:MTZIP},  (\ref{M8C1}) is the original fort cover constraint from Model \ref{Def:FortIP}; the rest of the constraints are the MTZ constraints. In particular, (\ref{M8C2}) ensures that there is an edge chosen from $\beta$ to some vertex that will be the root of the directed spanning tree of the zero forcing set; (\ref{M8C3}) ensures that each vertex has an incoming edge. Constraint (\ref{M8C4}) ensures that vertices connected to $\alpha$ cannot be used to connect to any other vertices; (\ref{M8C5}) and (\ref{M8C6}) ensure that there are no cycles in the chosen edges.  Constraint (\ref{M8C7}) ensures that vertices chosen to be in the forcing set must be in the spanning tree instead of connected to $\alpha$.  A solution of Model~\ref{Def:MTZIP} is a minimum connected forcing set by Theorem \ref{thm4} and by the fact that MTZ constraints impose connectivity of the selected set of vertices.

\begin{Model_float}
\begin{Model}{IP model for Connected Zero Forcing using MTZ constraints}
\label{Def:MTZIP}
\begin{eqnarray}
\nonumber\min && \sum_{\mathclap{v \in V}}s_{v}\\
\emph{s.t.:}&& \sum_{\mathclap{v \in B}}s_v \geq 1 \qquad \forall B \in \mathcal{B}\label{M8C1}\\
&&\sum_{\mathclap{v \in V}}y_{\beta,v} = 1\label{M8C2}\\
&&\sum_{\mathclap{i: (i,v) \in E}}y_{i,v} = 1 \qquad \forall v \in V\label{M8C3}\\
&&y_{\alpha,v} + y_{v,i} \leq 1\qquad \forall (v,i) \in E\label{M8C4}\\
&&(n+1)y_{i,v} + u_i - u_v + (n-1)y_{v,i} \leq n\qquad\forall (v,i) \in E\label{M8C5}\\
&&(n+1)y_{i,v} + u_i - u_v \leq n \qquad\forall (v,i) \in E_{new}\label{M8C6}\\
&& x_v = 1 - y_{\alpha,v} \qquad \forall v \in V\label{M8C7}\\
\nonumber&& y_{\alpha,\beta} = 1, u_{\alpha} = 0\\
\nonumber&&1 \leq u_v \leq n+1 \qquad \forall v \in V \cup \lbrace \beta \rbrace\\
\nonumber&& s\in \{0,1\}^n,y \in \{0,1\}^{m+2n+1}, u \in \mathbb{Z}^{n+2}
\end{eqnarray}
\end{Model}
\end{Model_float}

Rather than using additional variables, the second method for enforcing connectivity relies on adding valid inequalities which cut off disconnected solutions. These valid inequalities are known as $a$,$b$-separation inequalities; the idea behind them is that if a set $C\subset V$ is a vertex cut separating vertices $a$ and $b$ in a graph $G=(V,E)$, and both $a$ and $b$ are chosen to be in connected zero forcing set, then some vertex from $C$ must also belong to this forcing set.  Model \ref{Def:SeparationIP} gives the complete formulation for connected zero forcing using $a$,$b$-separation inequalities. Constraint (\ref{M9C1}) is the original fort cover constraint from Model \ref{Def:FortIP}, and (\ref{M9C2}) expresses the $a$,$b$-separation inequalities.

The $a$,$b$-separation inequalities can be separated efficiently using the observation that if the chosen zero forcing set $Z$ is not connected, then the set $ C = V\backslash Z$ must be a vertex cut separating at least two vertices $a \in Z$ and $b \in Z$.  However, as was pointed out by Buchanan et al. \cite{BuchananSangSungButenkoPasiliao}, the resulting vertex cuts in such an implementation are likely not minimal.  Since the decision variable for each vertex in $C$ appears in these constraints, the constraints are stronger when the size of the vertex cut $S$ is minimized. Therefore, vertices could be deleted from a vertex cut of $G$ until it becomes inclusion-minimal. For the dominating set problem (which was the focus of \cite{BuchananSangSungButenkoPasiliao}), a valid cutting plane can be obtained from a vertex cut; however, a zero forcing set does not have to be dominating.  Therefore, we also require that the vertex cut must be an $a$,$b$-separator for $a,b \in Z$.  Fischetti et al. \cite{FischettiConnectivity} give a method for finding $a$,$b$-separators between two components in a graph; we implemented and used a similar $a,b$-separator algorithm which gives a minimal $a,b$-separator (see also the algorithm of Buchanan et al. \cite{BuchananSangSungButenkoPasiliao} for minimal vertex cuts).

\begin{Model}{IP model for Connected Zero Forcing using $a$,$b$-separator constraints}
\label{Def:SeparationIP}
\begin{eqnarray}
\nonumber\min && \sum_{\mathclap{v \in V}}s_{v}\\
\emph{s.t.:} && \sum_{\mathclap{v \in B}}s_v \geq 1  \qquad \forall B \in \mathcal{B}\label{M9C1}\\
&& s_a + s_b - \sum_{\mathclap{v \in C}}s_v\leq 1 \qquad \forall a,b \in V \emph{ with } C \emph{ an } a,b\emph{-separator}\label{M9C2}\\
\nonumber &&s\in\{0,1\}^n
\end{eqnarray}
\end{Model}

%
%

\section{Computational Results}
\label{sec:Zero_Forcing_Results}
This section presents implementation details and computational results for finding minimum zero forcing sets and minimum connected forcing sets using the methods described thus far. In particular, for the zero forcing problem, we compare the 
\begin{itemize}
\setlength\itemsep{0em}
\item Wavefront algorithm (Algorithm \ref{algorithm_wavefront}), 
\item Infection model (Model \ref{Def:InfIP}), 
\item Fort Cover model without checking for facet-inducing forts (Model \ref{Def:FortIP} together with Model \ref{Def:FindFortIP}),
\item Fort Cover model with simplified checking for facet-inducing forts (Model~\ref{Def:FortIP} together with Model \ref{Def:FindFortIP} and the simplified version of Model \ref{Def:FacetIP}),
\item Extended Fort Cover model (Model \ref{Def:Fort_NeighborhoodIP} together with Model \ref{Def:FindMinimalBorderFortIP}), 
\item Maximal Closure model without checking for facet-inducing forts (Model \ref{Def:FortIP} with maximal closure method for generation of minimal forts),
\item Maximal Closure model with simplified checking for facet-inducing forts (Model \ref{Def:FortIP} with maximal closure method for generation of minimal forts and the simplified version of Model \ref{Def:FacetIP}).
\end{itemize}
For the connected forcing problem, we compare the
\begin{itemize}
\setlength\itemsep{0em}
\item Brute Force algorithm, 
\item Branch-and-Bound algorithm (Algorithm~\ref{alg:BandBZeroForcing}), 
\item Fort Cover model with MTZ constraints (Model \ref{Def:MTZIP} together with Model~\ref{Def:FindFortIP}),  
\item Fort Cover model with $a$,$b$-separator constraints (Model \ref{Def:SeparationIP} together with Model \ref{Def:FindFortIP}).
\end{itemize}
These methods are respectively labeled Wavefront, Infection, FC no facet, FC w/facet,  Ext. Cover, MC no facet, MC w/facet, Brute Force, B\&B, MTZ, and $a$,$b$-sep in table headings in the next section.

\subsection{Implementation Details}

Our computational results were obtained on a Dell PowerEdge R330 with an Intel(R) Xeon(R) CPU E3-1270 v5 @ 3.60GHz, 16 GB of RAM, and Red Hat Linux version 4.8.5-11.  Integer programs were solved using Gurobi version 7.5.2 set to use a single thread. The Brute Force, Branch-and-Bound, and Wavefront algorithms were implemented in C++ and compiled with g++ version 4.8.5. Implementations of all programs and models used can be obtained at \url{https://github.com/calebfast/zero_forcing}.

We tested the different solution approaches on several standard benchmark DIMACS10 \cite{dimacs_instances} (adjnoun, celegansneural, chesapeake, dolphins, football, jazz, karate, lesmis, and polbooks) and IEEE graphs \cite{ieee_instances} (14-Bus, 24-Bus, 30-Bus, 39-Bus, 57-Bus, 118-Bus, 300-Bus, and 96-RTS), as well as on three classes of random graphs: cubic graphs, WS$(5,0.3)$ graphs, and WS$(10,0.3)$ graphs.  We also tested the basic zero forcing methods on a series of star graphs with up to 101 vertices. We used our own C++ implementation to generate random cubic graphs, and we used the connected Watts-Strogatz graph generator from the NetworkX version 1.8.1 package in Python 2.7.6 \cite{networkX} to generate WS$(5,0.3)$ and WS$(10,0.3)$ graphs; the graphs generated are connected and simple, with no loops or multiple edges. For each family of graphs, we generated five random instances with $10i$ vertices for $i\in\{1,\ldots,10\}$; we omitted WS$(10,0.3)$ graphs with 10 vertices since those are just complete graphs.
We ran each algorithm for up to two hours. If an integer program could not solve an instance to optimality within two hours, we recorded the lower bound returned by Gurobi and the incumbent solution at the point of timeout (which is an upper bound on the optimal solution).

When solving Model~\ref{Def:FortIP}  without generation of facet-inducing forts, a maximal set of disjoint minimum size forts was added to the formulation using Model~\ref{Def:FindFortIP} before solving. Other violated fort constraints were added to the model using a MIPSOL callback, which is invoked by Gurobi whenever it finds a new integral incumbent solution.  When a violated fort is found, the MIPSOL callback adds that fort to the formulation as a lazy constraint (to enable lazy constraints, the ``PreCrush'' and ``LazyConstraints'' parameters of Gurobi were both set to 1). If no more violated forts are found, then Gurobi terminates with an optimal solution.

Similarly, when solving Model~\ref{Def:FortIP}  with generation of facet-inducing forts, if a generated fort $F$ is not facet inducing, then the valid cut associated with the forts that show $F$ is not facet inducing is added instead of $F$. Our computational results in Table \ref{table_times_zf} show that while checking for facet-inducing forts sometimes provides a small benefit in average running time and reduces the number of forts that must be generated, it is usually not effective enough to increase the size of the instances that can be solved within 2 hours.  Because checking for facet-inducing forts provided no consistent benefit, subsequent models using fort constraints do not check whether the generated forts are facet inducing. 

The Maximal Closure model with and without checking for facet-inducing forts, and Model~\ref{Def:Fort_NeighborhoodIP}, were solved similarly to Model~\ref{Def:FortIP}. In Model~\ref{Def:Fort_NeighborhoodIP}, violated minimum border forts were generated using Model~\ref{Def:FindMinimalBorderFortIP}, and a maximal set of disjoint forts was added to the formulation before solving. In addition, for each $v\in V$, the fort $V\backslash cl(N[v])$ was also added to the formulation before solving. 

Model~\ref{Def:MTZIP} was solved exactly like Model~\ref{Def:FortIP}, with the addition of the MTZ constraints. Violated forts are added to the formulation by the MIPSOL callback as lazy constraints; when no more violated forts are found, Gurobi terminates with an optimal solution.

In Model~\ref{Def:SeparationIP}, both fort constraints and $a$,$b$-separation inequalities are added to the model using a MIPSOL callback. This callback first generates a minimum size violated fort by solving Model~\ref{Def:FindFortIP}, and adds it to the formulation as a lazy constraint.
When no more violated forts are found, the callback checks whether the current solution is connected. If it is not connected, the $a,b$-separator algorithm finds a minimal separator contained in the separator given by the vertices outside the current solution; the corresponding $a$,$b$-separation inequality is then added to Model~\ref{Def:SeparationIP} as a lazy constraint. When no violated forts are found and the solution is connected, Gurobi terminates with an optimal solution.

For all these models, the parameters not mentioned in the discussion above were left to their defaults in Gurobi.  Some testing showed that tuning certain parameters (such as the branching direction (BranchDir), aggressiveness of cut generation (Cuts), or the focus of the solver (MIPFocus)) could improve performance on some specific instances, but not in general.  The branching strategy was also left to the Gurobi default.  

Tables \ref{table_times_zf} and \ref{table_times_zf2} give the average runtimes of the different zero forcing algorithms for the graphs tested; Table \ref{Table:AverageConnectedWS_10_0.3_ZFTimes} gives the average runtimes of the different connected forcing algorithms; Table \ref{throttling_table} gives the average runtimes of Model \ref{Def:InfIP} modified to run with a limited number of timesteps. The reported times reflect the time taken by Gurobi to optimize the relevant models; they include the time necessary for setting up the Gurobi models, but do not include the time necessary for data input.


\begin{table}[hbt!]
\renewrobustcmd{\bfseries}{\fontseries{b}\selectfont}
 \setlength{\tabcolsep}{3pt}
\singlespace
\scriptsize
\centering
 \caption{Average runtimes, in seconds, for zero forcing algorithms on different graphs. The reported runtimes and value of $Z(G)$ in each row are the average over the instances that were successfully solved. A number $[x]$ indicates that only $x$ of the five instances of the specified size were solved; `T' indicates that none of the instances were solved within 2 hours or within the available amount of memory (16 GB). In instances that could not be solved by the IP methods, $\{\ell/u\}$ denotes the lower bound $\ell$ and upper bound $u$ on $Z(G)$ at the point of timeout (averaged over all instances).  Bold text indicates the best performance or best bound for each set of instances.}
 \begin{tabular}{l|SS[detect-weight]S[detect-weight,group-digits=integer]S[detect-weight,group-digits=integer]S[detect-weight,group-digits=integer]S[detect-weight,group-digits=integer]S[detect-weight,group-digits=integer]SS}
&\;\;{$|V|$}&{$Z(G)$}&{Wavefront}\;\;&{Infection}\;&{FC w/facet} \;& {FC no facet} \;& {Ext. Cover} \;& {MC w/facet} \; & {MC no facet} \up \down \\ \hline
\multirow{10}{*}{\rotatebox[origin=c]{90}{Cubic graphs\down}}& \;\;10 & 3.8 &  \bfseries 0.0020 &  0.028 &  0.053 &  0.025 &  0.019 & 0.03 & 0.0059\up \\
&\;\;20 & 5.2 &  \bfseries 0.015 &  17.81 &  0.48 &  0.40 &  0.11 & 0.17 & 0.0383\\
&\;\;30 & 6.6 &  \bfseries 0.13 &57.00 \;\;[4]&  2.09 &  2.05 &  0.69 & 1.04 & 0.42\\
&\;\;40 & 8.8 &  \bfseries 1.95 &  3623.95 \hspace{-3.5pt}[3] &  15.20 &  17.94 &  6.10 & 13.49 & 17.60\\
&\;\;50 & 9.2 & \bfseries 6.69 &  3660.42\hspace{-3.5pt}[1] &  85.61 &  100.07 &  30.90 & 216.77 & 368.99\\
&\;\;60 & 11.4 &  \bfseries 156.83 & \{3.0/11.4\}  &  3302.15\hspace{-3.5pt}[2] &  4398.27\hspace{-3.5pt}[2] &  3138.83 & \{9.4/11.4\} & \{9.2/11.4\}\\
&\;\;70 & 12.0 &  \bfseries 370.50 & \{4.6/12.0\}  & \{9.8/12.0\}  & \{9.8/12.0\}  &  2852.27\hspace{-3.5pt}[3] & \{8.8/12.0\} & \{8.0/12.0\}\\
&\;\;80 & 12.8 &  \bfseries 1615.85\hspace{-3.5pt}[4] & \{6.2/13.6\}  & \{10.4/13.2\}  & \{10.2/13.2\}  &  5341.15\hspace{-3.5pt}[2] & \{9.2/13.2\} & \{8.4/13.2\}\\
&\;\;90 & 13.0 &  \bfseries 3101.29\hspace{-3.5pt}[2] & \{2.0/14.6\}  & \{8.8/13.6\}  & \{9.4/13.6\}  & \{9.4/13.6\}  & \{7.8/13.8\} & \{7.4/13.8\}\\
&\;\;100 & N/A&  T & \{3.4/16.6\} & \{8.6/15.8\}  & \{$\mathbf{9.6/15.4}$\}  & \{8.6/16.0\}  & \{8.0/15.4\} & \{7.8/15.8\}\down\\  \hline\hline
\multirow{10}{*}{\rotatebox[origin=c]{90}{WS$(5,0.3)$\down}}& \;\;10 & 4.4 &  \bfseries 0.00088 &  0.20 &  0.088 &  0.049 &  0.042 &  0.054 & 0.0073 \up \\
&\;\;20 & 6.2 &  \bfseries 0.013 &  23.66 &  0.89 &  0.65 &  0.66 & 0.47 & 0.063\\
&\;\;30 & 7.0 &  \bfseries 0.051 &  1989.99\hspace{-3.5pt}[4] &  5.90 &  5.32 &  4.88 & 2.53 & 0.89\\
&\;\;40 & 9.4 &  \bfseries 0.91 & \{5.2/9.4\}  &  49.28 &  49.00 &  38.52 & 48.09 & 44.91 \\
&\;\;50 & 10.8 &  \bfseries 7.20 & \{2.0/11.0\}  &  934.64 &  946.72 &  687.45 &  3279.41\hspace{-4pt}[3]& 3892.06\hspace{-4pt}[3]\\
&\;\;60 & 11.6 &  \bfseries 47.55 & 2958.70\hspace{-3.5pt}[1]&  1505.02\hspace{-3.5pt}[3] &  1485.19\hspace{-3.5pt}[3] &  594.42[3] & 262.58[1] & 330.19[1] \\
&\;\;70 & 14.0 &  \bfseries 474.31 & \{1.4/14.2\}  & \{10.0/14.0\}  & \{10.0/14.0\} & \{10.2/14.0\}  & \{8.6/14.0\}  &  \{8.0/14.0\}\\
&\;\;80 & 14.8 &  \bfseries 1441.92 & \{0.0/16.6\}  & \{9.2/15.4\}  & \{9.2/15.4\}  & \{9.8/14.8\}  & \{8.0/15.4\} &  \{7.8/15.6\}\\
&\;\;90 & 16.0 &  \bfseries 4537.77\hspace{-3.5pt}[1] & \{1.0/17.2\}  & \{8.0/18.2\}  & \{8.6/16.8\}  & \{7.6/20.0\} & \{8.0/17.2\} & \{7.4/16.8\} \\ 
&\;\;100 & N/A &  T & \{1.0/19.6\}  & \{6.8/53.8\}  & \{7.4/21.0\}  & \{6.2/86.6\} & \{$\mathbf{7.4/19.2}$\} & \{$\mathbf{7.2/19.0}$\} \down\\  \hline\hline
\multirow{10}{*}{\rotatebox[origin=c]{90}{WS$(10,0.3)$\down}}& \;\;20 & 12.0 &  \bfseries 0.0059 &  6913.10\hspace{-3.5pt}[1] &  13.97 &  5.90 &  6.97  & 15.75 & 0.97  \up \\
&\;\;30 & 15.4 &  \bfseries 0.076 & \{0.6/15.4\}  &  503.58 &  435.93 &  703.70 & 982.9 &  792.22 \\
&\;\;40 & 18.0 &  \bfseries 0.64 & \{0.0/18.0\}  & \{15.2/18.0\}  & \{15.2/18.0\}  & \{14.2/18.0\}  &\{13.4/18.0\}  & \{13.4/18.0\} \\
&\;\;50 & 21.8 &  \bfseries 6.44 & \{0.0/22.4\}  & \{13.0/21.8\}  & \{13.0/21.8\}  & \{12.2/21.8\}  & \{11.2/21.8\} & \{11.0/21.8\} \\
&\;\;60 & 24.6 &  \bfseries 45.00 & \{0.0/25.4\}  & \{9.2/24.6\}  & \{9.2/24.6\}  & \{10.0/25.2\}  & \{10.0/24.8\} &  \{10.0/24.8\}\\
&\;\;70 & 27.4 &  \bfseries 287.17 & \{0.0/31.2\}  & \{6.0/70.0\}  & \{6.2/70.0\}  & \{6.2/51.6\}  & \{9.6/27.8\} &  \{9.0/27.8\}\\
&\;\;80 & 31.2 &  \bfseries 2753.69 & \{0.0/33.6\}  & \{4.4/80.0\}  & \{5.2/80.0\}  & \{4.4/80.0\}  & \{8.4/31.6\} &  \{8.0/31.6\}\\
&\;\;90 & N/A &  T & \{0.0/37.4\}  & \{2.4/90.0\}  & \{4.4/90.0\} & \{3.8/90.0\}  & \{$\mathbf{8.2/35.6}$\} &  \{7.6/35.8\}\\
&\;\;100 & N/A &  T & \{0.0/41.8\}  & \{0.6/100.0\}  & \{4.6/100.0\}  & \{1.0/100.0\}  & \{$\mathbf{8.0/39.8}$\}  &  \{$\mathbf{8.0/39.8}$\} \down\\ \hline \hline

\multirow{10}{*}{\rotatebox[origin=c]{90}{Star graphs\down}}& 11&  9 & 0.30 & \bfseries 0.013 & 0.067 & 0.024  & 0.048  & 0.054 & 0.0084 \up\\
&21&  19 & 2234.53 & \bfseries 0.012 & 0.56 & 0.075  & 11.59  & 0.56 & 0.033 \\
&31&  29  & T  & \bfseries 0.025 & 1.99 & 0.23  & \;\;\;\{27/29\}  & 1.82 & 0.094 \\
&41&  39  & T  & \bfseries 0.039 & 4.84 & 0.56  & \;\;\;\{33/39\}  & 4.60 & 0.24 \\
&51& 49  & T  & \bfseries 0.063 & 10.59 & 1.13 & \;\;\;\{40/49\}  & 10.10 & 0.51 \\
&61&  59  & T  & \bfseries 0.14 & 20.64 & 1.92  & \;\;\;\{47/59\}  & 19.13 & 0.94 \\
&71&  69  & T  & \bfseries 0.20 & 33.87 & 3.46  & \;\;\;\{53/69\} & 33.11 & 1.64 \\
&81&  79  & T  & \bfseries 0.29 & 58.56 & 5.98  & \;\;\;\{65/79\} & 55.32 & 2.66 \\
&91&  89  & T  & \bfseries 0.47 & 86.91 & 7.75  & \;\;\;\{73/89\} & 81.26 & 3.00 \\
&101&  99  & T  & \bfseries 0.54 & 126.16 & 14.88  & \;\;\;\{81/99\}  & 117.29 &  5.20\\ 
 \end{tabular}
 \label{table_times_zf}
\end{table}

\begin{table}[hbt!]
\renewrobustcmd{\bfseries}{\fontseries{b}\selectfont}
 \setlength{\tabcolsep}{2pt}
\singlespace
\scriptsize
\centering
 \caption{Runtimes, in seconds, for zero forcing algorithms on real world graphs. `T' indicates that the instance was not solved within 2 hours or within the available amount of memory (16 GB). In instances that could not be solved by the IP methods, $\{\ell/u\}$ denotes the lower bound $\ell$ and upper bound $u$ on $Z(G)$ at the point of timeout.  Bold text indicates the best performance or best bound for each instance.}
 \begin{tabular}{l|SS[detect-weight]S[detect-weight,group-digits=integer]S[detect-weight,group-digits=integer]S[detect-weight,group-digits=integer]S[detect-weight,group-digits=integer]S[detect-weight,group-digits=integer]SS[detect-weight,group-digits=integer]S[detect-weight,group-digits=integer]S}
&\;{$|V|$}\;\;&{$Z(G)$}\;&{Wavefront}\;\;&{Infection}\;&{FC w/facet} \;& {FC no facet} \;& {Ext. Cover} \; & {MC w/facet} \;& {MC no facet} \;\down \\ \hline
karate& \;34& \;13 & 329.10& 6.00 & 0.42 & 0.59 & 0.39 & 2.31&\bfseries  0.16\up\\
chesapeake& \;39 & \;14 & \bfseries 10.91& \{4/14\}  & 49.70 & 37.36& 43.23 & 49.65& 35.43\\
dolphins& \;62& \;14 & 2405.56  & \{11/14\}  & \bfseries 246.24& 306.10 & 481.64 & 871.19 & 1335.88\\
lesmis& \;77& \;37  & T   & \{33/37\}  &  \bfseries2708.95& 5912.78& 4559.5 & \;\;\;\{36/37\} & \;\;\{36/37\}\\
polbooks& \;105 & N/A  & T   & \{2/28\}  &\{$\mathbf{14/27}$\}  &\;\{14/31\} & \;\{14/30\} & \;\;\;\{13/26\} & \;\;\{13/27\}\\
football&\;115 & N/A  & T   &  \{0/44\} & \{6/115\} &\;\{6/115\} & \;\{6/115\} & \;\;\;\{8/38\} & \;\;\{$\mathbf{9/38}$\}\\
celegansneural\;\,& \;297& N/A  & T   & \{21/133\}  & \{24/297\} & \;\{23/297\}& \;\{25/297\} & \;\;\;\{23/116\}& \;\;\{$\mathbf{28/100}$\}\\
adjnoun& \;112 & N/A  & T   & \{4/32\}  & \{9/37\} & \;\{9/112\}& \;\{7/55\} & \;\;\;\{9/32\}& \;\;\{$\mathbf{9/30}$\}\\
jazz& \;198 & N/A  & T   & \{0/104\} &  \{29/198\} & \;\{28/198\} & \;\{29/198\} & \;\;\;\{22/106\} & \;\;\{$\mathbf{27/96}$\}\\
IEEE\_14\_bus& \;14 & 4 & \bfseries0.0067  & 0.089 & 0.040& 0.029&  0.0088 & 0.053& 0.013\\
IEEE\_24\_bus& \;24 & 6 & 0.070 & 3.75 &  0.077 & 0.082&  0.038 &0.039 & \bfseries0.0061\\
IEEE\_30\_bus& \;30 & 7 & 0.82 & 0.44& 0.27& 0.25 & 0.15 & 0.12& \bfseries 0.033\\
IEEE\_39\_bus& \;39 & 7 & 6.69 & 0.15 & 0.22 &  0.24& 0.25 &0.18 & \bfseries0.043\\
IEEE\_57\_bus& \hspace{2.5pt}57 & 9 & 18.76 & 58.28 & 6.41& 5.79 & 5.68 & 2.83& \bfseries 1.97\\
IEEE\_RTS\_96& \;73 & \;15  & T   & \{12/15\}  & 6.36 &  5.63 & 7.47 & 2.53& \bfseries0.78\\
IEEE\_118\_bus& \;118& \;26  & T  & \bfseries 734.96& 5608.36& \;\{25/26\}   & \;\{25/26\} & \;\;\;\{24/26\}& \;\;\{24/26\}\\ 
IEEE\_300\_bus& \;300 & N/A  & T &\{$\mathbf{73/75}$\}  &\{68/75\} &  \;\{66/75\}  & \;\{66/76\} &\;\;\;\{66/75\} & \;\;\{66/75\}\\
 \end{tabular}
 \label{table_times_zf2}
\end{table}

\begin{table}[hbt!]
\renewrobustcmd{\bfseries}{\fontseries{b}\selectfont}
\centering
\singlespace
\scriptsize
 \caption{Average runtimes, in seconds, for connected forcing algorithms on different graphs. The reported runtimes and value of $Z_c(G)$ in each row are the average over the instances that were successfully solved. A number $[x]$ indicates that only $x$ of the five instances of the specified size were solved; `T' indicates that none of the instances were solved within 2 hours or within the available amount of memory (16 GB). In instances that could not be solved by the IP methods, $\{\ell/u\}$ denotes the lower bound $\ell$ and upper bound $u$ on $Z_c(G)$ at the point of timeout (averaged over all instances). For the B\&B method, $\{u\}$ denotes the upper bound on $Z_c(G)$ at the point of timeout.  Bold text indicates the best performance or best bound for each set of instances.}
 \setlength{\tabcolsep}{6pt}
 \begin{tabular}{l|SS[detect-weight,group-digits=integer]S[detect-weight,group-digits=integer]S[detect-weight,group-digits=integer]S[detect-weight,group-digits=integer]S[detect-weight,group-digits=integer]}
&{\hspace{-10pt} $|V|$} & {$Z_c(G)$}& {Brute Force} &{B\&B} & {MTZ} &{ $a$,$b$-sep}\up \down \\ \hline
& 10 & 3.8 &  0.0034 &  \bfseries 0.0018 &  0.038 &  0.018 \up \\
&20 & 5.4 & 0.37  &  \bfseries 0.011 &  0.16 &  0.10\\
&30 & 7.8 & 982.91  &  \bfseries 0.20 &  0.48 &  0.28\\
&40 & 9.8 & T  &  368.26 &  1.78 & \bfseries 0.87\\
Cubic graphs&50 & 10.4 & T  &  15.14 &  12.12 & \bfseries 2.87\\
&60 & 12.0 & T  &  377.66 &  40.16 &  \bfseries 9.86\\
&70 & 13.4 & T  &  199.92~\,[4] &  115.05 & \bfseries 39.04\\
&80 & 15.6 & T  &  3686.22[3] &  814.43 &  \bfseries 346.37\\
&90 & 15.2 & T  &  2229.52[4] &  1945.35 &  \bfseries 550.72\\
&100 & 17.4 & T  & \;\;\;\;\{51.4\}  &  5183.61[2] &  \bfseries 2002.71[4]\down \\ \hline\hline
& 10 & 4.4 &  0.0063 &  \bfseries 0.0051 &  0.058 &  0.044 \up \\
&20 & 6.2 & 1.26  &  \bfseries 0.14 &  0.38 &  0.21\\
&30 & 7.0 & 66.35  &  \bfseries 0.94&  2.52 &  1.97\\
&40 & 9.4 & 5972.53~[2]  &  17.93~~~[4] &  20.37 &  \bfseries 10.34\\
WS$(5,0.3)$&50 & 10.8 & T  &  754.94\,~[4] &  201.18 &  \bfseries 66.61\\
&60 & 12.0 & T  &  1535.009[2] &  2133.40[4] &  \bfseries 849.68\\
&70 & 13.5 & T  & \;\;\;\;\{48\}  &  4079.68[1] &  \bfseries 2492.52[2]\\
&80 & N/A & T  & \;\;\;\;\{41.6\} & \{10.6/15.8\}  &  \{$\mathbf{11.6/15.8}$\}\\
&90 & N/A & T  & \;\;\;\;\{47.2\}  & \{10.2/17.2\} & \{$\mathbf{11.0/17.0}$\} \\
&100 & N/A & T  & \;\;\;\;\{84.4\}  & \{$\mathbf{9.8/19.0}$\}  &  \{10.2/20.2\}\down\\ \hline \hline
& 20 & 12.0 &  83.86 &  11.6 &  14.3 &  \bfseries 7.52 \up \\
&30 & 15.4 & T  & 6493.70[1] &  2751.07[2] &  \bfseries 560.59\\
&40 & N/A & T  & \;\;\;\;\{18\}  & \{13.8/18.0\}  & \{$\mathbf{14.8/18.0}$\} \\
WS$(10,0.3)$&50 & N/A & T  & \;\;\;\;\{28.6\}  &  \{12.0/21.8\} & \{$\mathbf{13.0/21.8}$\} \\
&60 & N/A & T  & \;\;\;\;\{60\}  &  \{$\mathbf{10.2/24.8}$\} & \{9.4/24.6\} \\
&70 & N/A & T  & \;\;\;\;\{70\}  &  \{$\mathbf{8.0/32.2}$\} &  \{7.2/41.2\}\\
&80 & N/A & T  & \;\;\;\;\{80\}  &   \{$\mathbf{6.4/60.6}$\}&  \{6.2/72.8\}\\
&90 & N/A & T  & \;\;\;\;\{90\}  &   \{4.4/90.0\}&  \{$\mathbf{4.6/90.0}$\}\\
&100 & N/A & T  & \;\;\;\;\{100\}  &  \{$\mathbf{4.6/92.2}$\}&  \{4.6/100.0\}\down\\ \hline \hline
karate& 34& 14  & T   & \;\;\;\;\{14\}  & 2.03 & \bfseries 0.12\up\\
chesapeake& 39& 14  & T   & \;\;\;\;\{15\}   & \{13/14\}  & \bfseries 50.80\\
dolphins& 62& 18  & T   & \;\;\;\;\{62\}   & \{14/18\}  & \bfseries 4870.89\\
lesmis& 77& 40  & T   & \;\;\;\;\{77\}   & \{37/40\}  & \bfseries 812.36\\
polbooks& 105& N/A  & T   & \;\;\;\;\{105\}   & \{$\mathbf{14/28}$\}  & \{14/29\}\\
football& 115& N/A  & T   & \;\;\;\;\{115\}   & \{$\mathbf{6/115}$\}  & \{$\mathbf{6/115}$\}\\
celegansneural& 297& N/A  & T   & \;\;\;\;\{297\}   & \{23/297\}  & \{$\mathbf{36/297}$\}\\
adjnoun& 112& N/A  & T   & \;\;\;\;\{112\} & \{11/37\}  & \{$\mathbf{11/36}$\} \\
jazz& 198 & N/A  & T   & \;\;\;\;\{198\} & \{28/198\} & \{$\mathbf{28/197}$\}\\
IEEE\_14\_bus& 14& 4  & 0.01 & \bfseries 0.0096& 0.025&  0.032\\
IEEE\_24\_bus& 24& 7  & 19.60  &  0.11  & 0.13   & \bfseries 0.043 \\
IEEE\_30\_bus& 30& 9  & T  & 6.041& 0.33& \bfseries 0.059\\
IEEE\_39\_bus& 39& 15  & T  & 21.44& 0.72 & \bfseries 0.11\\
IEEE\_57\_bus& 57& 11  & T  & 2235.28& 1.97& \bfseries 0.55\\
IEEE\_RTS\_96& 73& 22  & T   & \;\;\;\;\{73\}  & 239.47& \bfseries 12.07\\
IEEE\_118\_bus& 118& 35  & T   & \;\;\;\;\{118\}   & \{32/35\}  & \bfseries 2459.81\\
IEEE\_300\_bus& 300& N/A  & T   & \;\;\;\;\{300\}   & \{90/118\}  & \{$\mathbf{102/117}$\}\\
 \end{tabular}
\label{Table:AverageConnectedWS_10_0.3_ZFTimes}
\end{table}

\begin{table}[hbt!]
\centering
\singlespace
\scriptsize
 \caption{Average running times, in seconds, for Model \ref{Def:InfIP} to reach an exact solution with different numbers of timesteps. The $\Delta Z$ columns give the average difference between $Z(G)$ and the minimum forcing set for the given number of timesteps. A number $[x]$ indicates that only $x$ of the five instances of the specified size were solved; in these cases, the reported result is the average time for the $x$ instances that were successfully solved. `T' indicates that none of the instances were solved within 2 hours or within the available amount of memory (16 GB).}
  \setlength{\tabcolsep}{3pt}
 \begin{tabular}{l|SS[detect-weight,group-digits=integer]S[detect-weight,group-digits=integer]S[detect-weight,group-digits=integer]S[detect-weight,group-digits=integer]S[detect-weight,group-digits=integer]S[detect-weight,group-digits=integer]S[detect-weight,group-digits=integer]S[detect-weight,group-digits=integer]}
\multicolumn{2}{c}{} & \multicolumn{2}{c}{$T=1$} & \multicolumn{2}{c}{$T=2$} & \multicolumn{2}{c}{$T=4$} & \multicolumn{2}{c}{$T=8$}\\
\cmidrule(r){3-4} \cmidrule(r){5-6} \cmidrule(r){7-8} \cmidrule(r){9-10}
&\;\;$|V|$&Time\;\;&$\Delta Z$\;&Time\;\;&$\Delta Z$\;&Time\;\;&$\Delta Z$\;&Time\;\;&$\Delta Z$\;\up \down \\ \hline
\multirow{10}{*}{\rotatebox[origin=c]{90}{Cubic graphs\down}}& 
10 &  0.012 & 2.4 &  0.025 & 1.0 &  0.028 & 0.0 &  0.032 & 0.0\up\\
&20 &  0.062 & 6.0 &  5.65 & 3.2 &  9.70 & 0.6 &  13.53 & 0.0\\
&30 &  0.18 & 10.2 &  111.49 & 6.2 &  1122.53 & 2.6 &  895.83 & 0.4\\
&40 &  1.39 & 13.8 &  3383.02[3] &  8.0& T & & T \\
&50 &  2.39 & 19.0 & T & & T & & T \\
&60 &  7.63 & 21.8 & T & & T & & T \\
&70 &  12.50 & 26.8 & T & & T & & T \\
&80 &  22.08 & 31.4  & T & & T & & T \\
&90 &  51.38 & N/A & T & & T & & T \\
&100 &  100.61 & N/A & T & & T & & T 
\down\\\hline\hline
\multirow{10}{*}{\rotatebox[origin=c]{90}{WS$(5,0.3)$\down}}& 
10 &  0.012 & 1.8 &  0.080 & 0.8 &  0.16 & 0.2 & 0.19 & 0.0\up\\
&20 &  0.064 & 6.4 &  10.74 & 3.8 &  32.60 & 1.6 & 18.97 & 0.0\\
&30 &  0.41 & 11.6 &  654.76 & 7.2 &  2598.63[2] & 3.5 & T &\\
&40 &  1.74 & 14.8 & T  & &T  & &T \\
&50 &  4.74 & 19.8 & T  && T  && T \\
&60 &  10.46& 25.4 & T  & &T  && T \\
&70 &  25.74 & 28.0 & T  & &T && T \\
&80 &  88.43 & N/A & T  & &T & & T \\
&90 &  138.37 & N/A & T  && T & & T \\
&100 &  191.57 & N/A & T  & &T & & T 
\down\\  \hline\hline
\multirow{7}{*}{\rotatebox[origin=c]{90}{WS$(10,0.3)$\down}}& 
20 &  0.70 & 3.0 &  240.66 & 1.2 &  1112.63 & 0.0 &  4252.58[4] & 0.0\up\\
&30 &  9.25 & 6.4 & T  & &T & & T \\
&40 &  66.53 & 11.0 & T  & &T  && T \\
&50 &  388.04 & 14.2 & T & & T & & T \\
&60 &  816.90 & 18.2 & T & & T  && T \\
&70 &  1617.63[4] & 21.75  &T & & T  & & T \\
&80 &  3863.66 & 25.4 & T & & T  && T 
\down\\
 \end{tabular}
 \label{throttling_table}
\end{table}

\subsection{Comparison and Discussion}

The computational results in Tables \ref{table_times_zf} and \ref{table_times_zf2} show that for the zero forcing problem, the Wavefront algorithm performs best for random cubic graphs and Watts-Strogatz graphs, while the integer programming models are generally faster for real-world graphs corresponding to electrical power grids and other networks. At the bottom of Table~\ref{table_times_zf}, we show a case when the IP models perform much better than Wavefront. Similar behavior can be observed in graphs where small subsets of vertices do not force many vertices outside of the subsets; this is the case in some of the IEEE graphs for which Wavefront also has inferior performance. The Wavefront algorithm is also not as easily adapted as the integer programming methods to additional constraints, such as ensuring connectivity or limiting the number of timesteps used.

For random graphs, the Wavefront algorithm is fastest, followed by the Extended Fort Cover model, the standard Fort Cover models, and the Maximal Closure models; all these approaches considerably outperform the Infection model. Facet-inducing constraints usually provide a slight speedup in the Fort Cover and Maximal Closure models, but do not allow any instances to be solved which could not be solved without facet-inducing constraints. All five Fort Cover models can handle similarly sized random cubic and Watts-Strogatz graphs, although the Extended Fort Cover model solves the instances faster on average, especially in sparser graphs. 

For small graphs, using minimal fort constraints with the Maximal Closure models generally gives better performance than using minimum fort constraints generated by an IP. This can be explained by the fact that in small graphs it is more likely that a minimal fort is also minimum --- hence, we get the benefit of a minimum fort without the cost of solving an IP to find it. However, for larger graphs, minimum forts generally perform better than minimal forts. The Maximal Closure models (with and without facets) do not solve any instances that the other methods cannot solve, but they fail to solve some instances that the other methods can solve; due to this, the Maximal Closure models are omitted from Table \ref{Table:AverageConnectedWS_10_0.3_ZFTimes}.

The differences in runtimes for the different types of graphs indicate that the integer programs are sensitive to the density and vertex degrees of the graphs, while the Wavefront algorithm is less sensitive to these changes, and is primarily affected by size of the graphs. This can be seen by comparing runtimes for cubic graphs with runtimes for Watts-Strogatz graphs, or runtimes for DIMACS graphs with runtimes for IEEE graphs. For example, the DIMACS football instance has roughly the same number of vertices and three times as many edges as the IEEE 118-Bus instance; the latter is solved or closely  bounded by the IP models, while the former is not solved by any models, and the bounds have a very wide gap. In Table \ref{table_times_zf2}, for the graphs corresponding to electrical power grids and other real-world networks, the Wavefront algorithm was generally outperformed by the integer programming models, which were mutually competitive in performance. 

For the connected forcing problem, the computational results in Table \ref{Table:AverageConnectedWS_10_0.3_ZFTimes} show that the Branch-and-Bound algorithm performs best on small, sparse graphs, but is outperformed by the integer programming models as the size and density of the graphs increases. This is because the Branch-and-Bound algorithm relies on enumerating connected induced subgraphs, and larger, denser graphs have significantly more such subgraphs. The Fort Cover model with $a,b$-separation constraints solves the largest instances out of any method. This result is in line with other results from the literature, as the $a,b$-separation constraints have outperformed the MTZ constraints for other problems such as connected domination \cite{BuchananSangSungButenkoPasiliao}. For all of the DIMACS10 graphs, the Fort Cover Model with $a$,$b$-separation constraints was significantly faster than the model with MTZ constraints and the Branch-and-Bound algorithm; for most other graphs, the three methods were able to handle roughly the same sized graphs. All three of the nontrivial approaches are faster and able to handle larger graphs than the Brute Force approach.

When comparing Models \ref{Def:MTZIP} and \ref{Def:SeparationIP}, we see that the MTZ constraints are able to solve similar sizes of instances as the $a,b$-separation constraints, although the $a,b$-separation constraints give faster runtimes.  Note that as the average degree of the vertices increases, the likelihood that a chosen subset of vertices will induce a connected graph increases. Therefore, for graphs with high average degree, the $a$,$b$-separation inequalities were usually not necessary, and the model was solved as a basic zero forcing problem. 

When considering the bounds for the IP models for large unsolved instances, we see that the Maximal Closure methods generally give the best bounds, followed by the Infection model. For connected forcing, the bounds given by the MTZ constraints and the $a,b$-separator constraints are roughly the same. In all cases, the graph density appears to affect the quality of the bounds: the gaps between the upper and lower bounds are smaller for unsolved instances of sparse graphs. In some cases, especially in sparser graphs, the gap between the upper and lower bounds is quite small; for example, in the IEEE 300-Bus graph, the Infection Model had a gap of 2 between the upper and lower bound. Thus, the IP models could sometimes be used to accurately approximate the zero forcing numbers of graphs which are too large for exact computation. However, in large graphs which are denser, such as the DIMACS celegansneural or jazz instances, the gaps between the upper and lower bounds are very large. The combinatorial algorithms we considered ---  Wavefront, the Brute Force algorithm, and the Branch-and-Bound algorithm --- only lend one nontrivial bound (lower bound in the case of Wavefront and brute force, and upper bound in the case of the Branch-and-Bound method) at the point of timeout. As such, they are not as useful as the integer programming models in providing heuristic solutions for large instances. For example, as can be seen from Table~\ref{Table:AverageConnectedWS_10_0.3_ZFTimes}, the upper bound given by the Branch-and-Bound algorithm was usually equal $n$, and was far from the true solution.

Finally, Table \ref{throttling_table} shows computational results for the problem of zero forcing with a bounded number of timesteps.  Model \ref{Def:InfIP} runs faster and is able to handle larger graphs when it is coupled with lower bounds on the number of timesteps $T$. For small values of $T$, the model solved for all graphs, although it was still somewhat impaired by the graph density. As seen in the $\Delta Z$ columns of Table \ref{throttling_table}, the sizes of the zero forcing sets with bounded number of timesteps approach the sizes of the minimum zero forcing sets as $T$ grows. However, the increase in $T$ also causes increased runtime due to the big-$M$ constraints in the model. 

\section{Conclusion and Future Work}

\label{sec:ZeroComputationConclusions}

This paper introduced new methods for computing the zero forcing and connected forcing numbers of graphs. We presented combinatorial algorithms, as well as integer programming formulations based on an infection perspective and a set-covering perspective of zero forcing. We explored several solution strategies for these models, drawing from different areas of integer programming and polyhedral theory, and we compared their performance on random cubic graphs, Watts-Strogatz graphs, and various standard benchmark graphs. Our computational experiments show that the Wavefront algorithm generally outperforms the integer programming models for zero forcing on random cubic graphs and Watts-Strogatz graphs, while the integer programming models are faster for real-world graphs corresponding to electrical power grids and other networks. Moreover, our algorithms for connected forcing were comparable in performance to Wavefront and the zero forcing models (and in some cases they were faster and able to handle slightly larger graphs). We also presented an integer program for finding a set of minimum cardinality which forces a graph within a fixed number of timesteps; this model performed very well for small numbers of timesteps. It would be interesting to extend the Fort Cover models to solve this fixed-timestep problem by adding certain valid inequalities, and compare them against the modified Infection model. It would also be interesting to experiment by varying the implementation of the IP models, e.g., adding a violated cut for every pair of disconnected vertices when using Model \ref{Def:SeparationIP}, or adding violated $a,b$-separation cuts in each callback rather than when the callback solution satisfies all fort inequalities.  Our preliminary tests showed that these variations do not seem to provide a benefit, but they may be beneficial for graph families that were not tested.   

Some of the difficulty in solving the zero forcing problem is due to the symmetry of solutions; this symmetry arises from the fact that for each zero forcing set and any associated set of forcing chains, another zero forcing set of equal size can be obtained by choosing the terminals of the forcing chains \cite{Barioli}. Such sets of vertices are nearly indistinguishable in many of the algorithms and formulations, and this symmetry is harder to detect than simple isomporphisms in the graph. Any method for dealing with the symmetry of zero forcing has the potential to drastically improve the performance of the integer programs presented in this paper. Therefore, a direction for future work is to focus on breaking this symmetry. Note that this symmetry is somewhat less prevalent in connected forcing, since the set of terminals of forcing chains associated with a connected forcing set is not always connected; this may be part of the reason why the connected variants of the integer programs sometimes performed better.


\section*{Acknowledgements}
We thank the five anonymous referees whose helpful and constructive comments greatly improved the presentation and results of the paper. This work was supported by the National Science Foundation, grant numbers 1450681, CMMI-1300477, and CMMI-1404864.


\begin{thebibliography}{99}
\singlespace

\bibitem{aazami_pd}
A. Aazami.
Domination in graphs with bounded propagation: algorithms, formulations and hardness results.
\emph{Journal of Combinatorial Optimization}, 19(4):
429--456, 2010.


\bibitem{aazami}
A. Aazami.
Hardness results and approximation algorithms for some problems on graphs.
PhD thesis, University of Waterloo, 2008.

\bibitem{target1}
E. Ackerman, O. Ben-Zwi, and G. Wolfovitz.
Combinatorial model and bounds for target set selection. 
\emph{Theoretical Computer Science}, 411(44-46):
4017--4022, 2010.


\bibitem{AIM-Workshop}
AIM Special Work Group.
Zero forcing sets and the minimum rank of graphs.
\emph{Linear Algebra and its Applications}, 428(7): 
1628--1648, 2008.

\bibitem{kforcing1}
D. Amos, Y. Caro, R. Davila, and R. Pepper. 
Upper bounds on the $k$-forcing number of a graph.
\emph{Discrete Applied Mathematics}, 181:
1--10, 2015.


\bibitem{AVELLA_BOCCIA_VASILYEV}
P. Avella, M. Boccia, and I. Vasilyev.
Computational experience with general cutting planes for the set covering problem. 
\emph{Operations Research Letters}, 37: 
16--20, 2009.
  
\bibitem{AVIS_FUKUDA}
D. Avis and K. Fukuda. 
Reverse search for enumeration.
\emph{Discrete Applied Mathematics}, 65:
21--46, 1996.


\bibitem{dimacs_instances}
D.A. Bader, A. Kappes, H. Meyerhenke, P. Sanders, C. Schulz and D. Wagner. Benchmarking for graph clustering and partitioning. \emph{Encyclopedia of Social Network Analysis and Mining}, 
73--82, 2014. Available at \url{https://www.cc.gatech.edu/dimacs10/downloads.shtml}.


\bibitem{BALAS_NG}
E. Balas and S.M. Ng. 
On the set covering polytope: I. All the facets with coefficients in $\{0,1,2\}$.
\emph{Mathematical Programming}, 43:
57--69, 1989.
 
\bibitem{zf_tw}
F. Barioli,  W. Barrett, S.M. Fallat, T. Hall, L. Hogben, B. Shader, P. van den Driessche, and H. van der Holst.
Parameters related to tree-width, zero forcing, and maximum nullity of a graph.
\emph{Journal of Graph Theory}, 72(2): 
146--177, 2013.


\bibitem{Barioli}
F. Barioli, W. Barrett, S. Fallat, H.T. Hall, L. Hogben, B. Shader, P. van den Driessche, and 
H. van der Holst. 
Zero forcing parameters and minimum rank problems. 
\emph{Linear Algebra and its Applications}, 433(2): 
401--411, 2010.




\bibitem{benson}
K. Benson, D. Ferrero, M. Flagg, V. Furst, L. Hogben, V. Vasilevska, and B. Wissman.
Power domination and zero forcing.
\emph{arXiv}:1510.02421, 2015.

\bibitem{target3}
O. Ben-Zwi, D. Hermelin, D. Lokshtanov, and I. Newman.
Treewidth governs the complexity of target set selection. 
\emph{Discrete Optimization}, 8(1):
87--96, 2011.


\bibitem{proptime_oriented}
A. Berliner, C. Bozeman, S. Butler, M. Catral, L. Hogben, B. Kroschel, J.C-H. Lin, N. Warnberg, M. Young. 
Zero forcing propagation time on oriented graphs. 
\emph{Submitted}, 2015.


\bibitem{brimkov2}
B. Brimkov and I.V. Hicks. 
Complexity and computation of connected zero forcing.
\emph{Discrete Applied Mathematics} 229: 
31--45, 2017.


\bibitem{brimkov}
B. Brimkov and R. Davila. 
Characterizations of the connected forcing number of a graph.
\emph{arXiv}:1604.00740, 2016.

\bibitem{brimkov3}
B. Brimkov, C.C. Fast, I.V. Hicks.
Graphs with extremal connected forcing numbers.
\emph{arXiv}:1701.08500, 2017.

\bibitem{brimkov_pd}
B. Brimkov, D. Mikesell, and L. Smith.
Connected power domination in graphs. 
\emph{arXiv}:1712.02388, 2017.

\bibitem{BuchananSangSungButenkoPasiliao}
A. Buchanan, J.S. Sung, S. Butenko, and E.L. Pasiliao. 
An integer programming approach for fault-tolerant connected dominating sets.
\emph{INFORMS Journal on Computing}, 27(1):
178--188, 2015.

\bibitem{quantum1}
D. Burgarth and V. Giovannetti.
Full control by locally induced relaxation.
\emph{Physical Review Letters}, 99(10): 
100501, 2007.

\bibitem{throttling}
S. Butler, and M. Young. 
Throttling zero forcing propagation speed on graphs.
\emph{Australasian Journal of Combinatorics}, 57: 
65--71, 2013.


\bibitem{logic1}
D. Burgarth, V. Giovannetti,  L. Hogben, S. Severini,  and M. Young.
Logic circuits from zero forcing.
\emph{arXiv}:1106.4403, 2011.

\bibitem{WavefrontAlgorithm}
S. Butler, L. DeLoss, J. Grout, H.T. Hall, J. LaGrange, T. McKay, J. Smith, and G. Tims.
Minimum Rank Library ({\textit{Sage}} programs for
  calculating bounds on the minimum rank of a graph, and for computing zero
  forcing parameters), 2014.
Available at \texttt{https://github.com/jasongrout/minimum{\_}rank}.

\bibitem{butler}
S. Butler, J. Grout, and H.T. Hall.
Using variants of zero forcing to bound the inertia set of a graph.
\emph{Electronic Journal of Linear Algebra}, 30: 2015.

\bibitem{Caro}
Y. Caro, D.B. West, and R. Yuster. 
Connected domination and spanning trees with many leaves.
\emph{SIAM Journal on Discrete Mathematics}, 13: 
202--211, 2000.

\bibitem{CarvajalConnectivity}
R. Carvajal, M. Constantino, M. Goycoolea, J.P. Vielma, and A. Weintraub.
Imposing connectivity constraints in forest planning models.
\emph{Operations Research}, 61(4):
824--836, 2013.

\bibitem{target2}
C.Y. Chiang, L.H. Huang, B.J. Li, J. Wu, and H.G. Yeh.
Some results on the target set selection problem. 
\emph{Journal of Combinatorial Optimization}, 25(4):
702--715, 2013.

\bibitem{iteration_index}
K.B. Chilakamarri, N. Dean, C.X. Kang, and E. Yi.
Iteration index of a zero forcing set in a graph. 
\emph{arXiv}:1105.1492, 2011.

\bibitem{networkX}
Connected Watts-Strogatz small-world graphs. NetworkX Documentation, 2013. Available at: \url{https://networkx.github.io/documentation/networkx-1.8.1/reference/generated/networkx.generators.random_graphs.connected_watts_strogatz_graph.html}.


\bibitem{DantFulkJohn}
G. Dantzig, R. Fulkerson, and S. Johnson.
Solution of a large-scale traveling-salesman problem.
\emph{Operations Research}, 2 
393--410, 1954.

\bibitem{Desormeaux}
W.J. Desormeaux, T.W. Haynes, and M.A. Henning. 
Bounds on the connected domination number of a graph.
\emph{Discrete Applied Mathematics}, 161(18): 
2925--2931, 2013.

\bibitem{Edholm}
C. Edholm, L. Hogben, J. LaGrange, and D. Row.
Vertex and edge spread of zero forcing number, maximum nullity, and minimum rank of a graph.
\emph{Linear Algebra and its Applications}, 436(12): 
4352--4372, 2012.


\bibitem{positive_zf2}
J. Ekstrand, et al. 
Positive semidefinite zero forcing. 
\emph{Linear Algebra and its Applications}, 439(7):
1862--1874, 2013.

\bibitem{Eroh}
L. Eroh, C. Kang, and E. Yi.
Metric dimension and zero forcing number of two families of line graphs.
\emph{arXiv}:1207.6127, 2012.

\bibitem{fallat}
S. Fallat and L. Hogben.
 The minimum rank of symmetric matrices described by a graph: A survey. 
\emph{Linear Algebra and its Applications}, 426: 
558--582, 2007.

\bibitem{FanWatson}
N. Fan and J.-P. Watson. 
Solving the connected dominating set problem and power dominating set problem by integer programming. 
\emph{International Conference on Combinatorial Optimization and Applications}, 
Springer Berlin Heidelberg,
pp. 371--383, 2012.

\bibitem{ForcingTheoryPaper}
C.C. Fast and I.V. Hicks.
The effect of vertex degrees on the zero-forcing number and iteration index of a graph.
\emph{Submitted}, 2016.

\bibitem{FischettiConnectivity}
M. Fischetti, M. Leitner, I. Ljubi\'{c}, M. Luipersbeck, M. Monaci, M. Resch, D. Salvagnin, and M. Sinnl.
Thinning out Steiner trees: a node-based model for uniform edge costs.
\emph{Mathematical Programming Computation},
1--27, 2016.
  
\bibitem{FISCHETTI_LODI}
M. Fischetti and A. Lodi.
Optimizing over the first Chv\'atal closure.
{\em Mathematical Programming}, 110:
3--20,  2007.


\bibitem{connected_dom}
F.V. Fomin, F. Grandoni, and D. Kratsch. 
Solving connected dominating set faster than $2^n$. 
\emph{Algorithmica} 52(2): 
153--166, 2008. 
 
\bibitem{signed_zf}
F. Goldberg and A. Berman. 
Zero forcing for sign patterns. 
\emph{Linear Algebra and its Applications}, 447: 
56--67, 2014.

\bibitem{powerdom3}
T. Haynes,  S. Hedetniemi, S. Hedetniemi, and M. Henning.
Domination in graphs applied to electric power networks.
\emph{SIAM Journal on Discrete Mathematics}, 15(4): 
519--529, 2002.


\bibitem{proptime1}
L. Hogben, N. Kingsley, S. Meyer, S. Walker, and M. Young. 
Propagation time for zero forcing on a graph.
\emph{Discrete Applied Mathematics}, 160(13):
1994--2005, 2012.

\bibitem{fractional_zf}
L. Hogben, K. F. Palmowski, D. E. Roberson, and M. Young. 
Fractional zero forcing via three-color forcing games.
\emph{Discrete Applied Mathematics}, 213:
114-129, 2016.


\bibitem{Huang}
L.-H. Huang, G. J. Chang, and H.-G. Yeh. 
On minimum rank and zero forcing sets of a graph.
\emph{Linear Algebra and its Applications}, 432: 2961--2973, 2010

\bibitem{ieee_instances}
Illinois Center for a Smarter Electric Grid, Power Flow Test Cases, 2018. Available at \url{http://icseg.iti.illinois.edu/power-cases/}.

\bibitem{pmu1}
K.G. Khajeh, E. Bashar, A.M. Rad, and G.B. Gharehpetian.
Integrated model considering effects of zero injection buses and conventional measurements on optimal PMU placement. 
\emph{IEEE Transactions on Smart Grid}, 8(2):
1006--1013, 2017.

\bibitem{kforcing2}
L. Lu, B. Wu, and Z. Tang. 
Proof of a conjecture on the zero forcing number of a graph.
\emph{Discrete Applied Mathematics}, 213:
223-237, 2016.

\bibitem{pmu2}
S.M. Mahaei and M.T. Hagh.  
Minimizing the number of PMUs and their optimal placement in power systems.
\emph{Electric Power Systems Research}, 83(1): 
66--72, 2012.

\bibitem{Meyer}
S. Meyer.
Zero forcing sets and bipartite circulants.
\emph{Linear Algebra and its Applications}, 436(4): 
888--900, 2012.

\bibitem{MillerTuckerZemlin}
C.E. Miller, A.W. Tucker, and R.A. Zemlin. 
Integer programming formulation of traveling salesman problems.
\emph{ Journal of the ACM}, 7:
326--329, 1960.

\bibitem{NemhauserWolsey}
G.L. Nemhauser and L.A. Wolsey.
\emph{Integer and Combinatorial Optimization}.
Wiley, New York, 1999.


\bibitem{QuintaoDaCunhaMateusLucena}
F.P. Quint\={a}o, A.S. da~Cunha, G.R. Mateus, and A. Lucena.
The $k$-cardinality tree problem: reformulations and Lagrangian relaxation.
\emph{Discrete Applied Mathematics}, 158(12):
1305 -- 1314, 2010.

\bibitem{row}
D.D. Row. 
A technique for computing the zero forcing number of a graph with a cut-vertex.
\emph{Linear Algebra and its Applications}, 436: 
4423--4432, 2012.

\bibitem{zf_np}
M. Trefois and J.C. Delvenne.
Zero forcing number, constrained matchings and strong structural controllability.
\emph{arXiv}:1405.6222v2, 2015.

\bibitem{WangBuchananButenko}
Y. Wang, A. Buchanan, and S. Butenko. 
On imposing connectivity constraints in integer programs.
\emph{Mathematical Programming}, 166: 1--31, 2017.

\bibitem{proptime2}
N. Warnberg.
Positive semidefinite propagation time.
\emph{Discrete Applied Mathematics}, 198:
274--290, 2016.

\bibitem{WattsStrogatz}
D.J. Watts and S.H. Strogatz.
Collective dynamics of `small-world' networks.
\emph{Nature}, 393:
440--442, 1998.


\bibitem{west}
D.B. West. 
\emph{Introduction to Graph Theory}. 
Prentice Hall, Inc., Upper Saddle River, NJ, 2001.


\bibitem{fast_mixed_search}
B. Yang. 
Fast-mixed searching and related problems on graphs. 
\emph{Theoretical Computer Science}, 507: 
100--113, 2013.

\bibitem{powerdom2}
M. Zhao,  L. Kang, and G. Chang.
Power domination in graphs.
\emph{Discrete Mathematics}, 306(15): 
1812--1816, 2006.

\end{thebibliography}
\end{document}